%% file: main.tex
\documentclass[aps,pra,reprint,groupedaddress,longbibliography]{revtex4-2}
\usepackage[T1]{fontenc}
\usepackage{lmodern}
\usepackage{amsmath,amssymb,amsthm,mathtools,bm,graphicx,booktabs,xcolor}
\usepackage{capt-of}
\usepackage{hyperref}
\newcommand{\suppsec}[1]{\ref{sec:supp:#1}}
\hypersetup{colorlinks=true,citecolor=blue!55!black,linkcolor=blue!55!black,urlcolor=blue!55!black}
\newtheorem{theorem}{Theorem}
\newtheorem{proposition}{Proposition}
\newtheorem{lemma}{Lemma}
\newcommand{\Tr}{\operatorname{Tr}}
\newcommand{\id}{\mathbb I}
\newcommand{\Vrec}{V_{\mathrm{rec}}}
\newcommand{\Bisep}{\operatorname{Bisep}}
\newcommand{\Ppar}{\Pi_{\mathrm{par}}}
\newcommand{\Pcorr}{\Pi_{\mathrm{corr}}}

\begin{document}
\title{Entanglement inference from quantum erasure and path discrimination}
\author{Mohd Asad Siddiqui}
\email{asad@ctp-jamia.res.in}
\affiliation{Department of Physics, Government Girls P.G. College, Sawai Madhopur, Rajasthan 322001, India}
\begin{abstract}
Fringes recovered by conditioning on quantum-eraser outcomes do not, by themselves, establish entanglement. We determine the minimum entanglement compatible with recorded eraser data when the preparation may be mixed, distinguishing these data from independently determined optimal which-path distinguishabilities and additional joint measurement outcomes. For noisy Greenberger-Horne-Zeilinger (GHZ) preparations, explicit compatible states establish sharp entanglement-depth thresholds and show how parity information strengthens certification. A matching converse establishes the exact threshold for two complete Pauli-outcome distributions. For an ideal three-qubit eraser, we classify the additional marker measurement directions that certify genuine multipartite entanglement within a fixed eraser configuration and determine the separate minima of compatible concurrence and squared GHZ fidelity for specified tilted axes. For noisy three-qubit data, an exact semidefinite characterization and constructive converse determine biseparable compatibility at arbitrary fixed local measurement axes, with rational certificates bounding one example. These results establish what entanglement is required by the measurement record, rather than inferred from assumptions about the source.
\end{abstract}
\maketitle

\section{Introduction}

In a two-path interferometer, a which-path marker can suppress the fringes observed when the marker outcomes are ignored. A quantum eraser measures the marker in another basis and sorts the detection events into subensembles: fringes and antifringes may then appear even though they cancel in the unsorted data \cite{Scully1982,Kwiat1992,Kim2000,EnglertBergou2000,Ma2013}. The recovered visibility characterizes the selected subensembles, not the unconditional interference pattern. Whether the same measurement records establish entanglement between the path and the markers is a separate question.

Recovered fringes alone cannot settle this question. Gogo, Snyder, and Beck showed that classically correlated preparations reproduce several, though not all, signatures of quantum erasure \cite{Gogo2005}; quantum-eraser protocols designed to verify electron--photon entanglement have also been proposed \cite{Henke2025}. Rather than assuming the intended preparation, we consider all states compatible with the observed records. These include the probabilities of the marker outcomes, the corresponding path populations, and both quadratures of the conditional interference fringe. We ask which separability classes remain compatible with these data.

The standard wave--particle duality bound relates unconditional visibility to optimal which-way distinguishability for a given state \cite{Englert1996}. Extensions to multipath interference, quantum coherence, and nonlocal settings have broadened the description of complementarity \cite{SiddiquiQureshi2015,Bera2015,SiddiquiQureshi2016,QureshiSiddiqui2017,SiddiquiQureshi2021}. Further connections involve predictability, entanglement, and decoherence \cite{Qureshi2021,Roy2022Triality,Sharma2020Decoherence,Ding2025}. Quantum erasure itself has been described through a duality--entanglement relation that accounts for partial recovery of interference \cite{Maleki2021}. These results characterize relations among interference, path information, and entanglement. Here the task is different: to determine the least entanglement required by specified measurement outcomes when the preparation may be mixed.

With several markers, which subsystems are measured matters as much as the choice of basis. Ghost-interference, three-slit, cavity, and Bell-state eraser schemes illustrate different ways of observing conditional interference and correlations \cite{ChingangbamQureshi2012,ShahQureshi2017,Rossi2013,Glick2017}. Which-path interactions can also generate many-body quantum correlations \cite{Ruimy2024}. In this setting, entanglement across the path--marker cut, pairwise entanglement, and genuine multipartite entanglement must be distinguished. Another distinction concerns the available path information: the optimal probability of correctly identifying a path is a single number, whereas the full joint outcomes of a measurement attaining that optimum contain more information.

Entanglement inference from incomplete measurement data is an established optimization problem \cite{AudenaertPlenio2006,WunderlichPlenio2009}. Stabilizer witnesses can require only a few local measurement settings \cite{TothGuhne2005,TothGuhnePRA2005,Zhao2019}, and suitably chosen observables can identify a Greenberger--Horne--Zeilinger (GHZ) state \cite{Yan2011}. Positive-partial-transpose-mixture criteria provide tests of genuine multipartite entanglement \cite{Jungnitsch2011}. More recent methods construct witnesses from restricted sets of observables \cite{Seong2026,Li2026Restricted}, while general numerical frameworks address multipartite entanglement certification \cite{Liu2026}. Measurement misalignment and limited access to correlations introduce further practical constraints \cite{Cao2024,Wu2026}. Our setting imposes a more specific constraint: the eraser measurement is fixed in advance. We determine what can be certified from its records and how additional local measurements can rule out separable or biseparable alternatives.

We begin with preparations that produce the same eraser fringes and optimal path distinguishabilities but have different entanglement depths. We then show how recorded parity and complete Pauli-outcome distributions strengthen certification. For a fixed three-qubit eraser, we classify the additional marker directions that certify genuine multipartite entanglement and determine the separate minima of compatible concurrence and squared GHZ fidelity for specified tilted axes. Finally, an exact semidefinite program (SDP) and constructive converse determine the white-noise biseparable-compatibility boundary at fixed local measurement axes, with rational certificates for one example. Together, these results distinguish interference recovered by conditioning from entanglement necessarily established by the recorded data.

\section{Eraser data and compatible states}

Let $A$ be a balanced path qubit, $\langle a|\rho_A|a\rangle=1/2$ for $a=0,1$, and let $M=M_1\cdots M_m$ comprise $m\ge2$ marker qubits. For three-qubit examples, we write $B=M_1$ and $C=M_2$. Operators $X_j,Y_j,Z_j$ are the Pauli matrices on party $j$; an omitted tensor factor is the identity. We write $\Bisep$ for the set of biseparable states, which are mixtures of states product across some bipartition. A state outside this set is genuinely multipartite entangled (GME). The entanglement depth is the least $k$ for which a state is a mixture of states product across groups of at most $k$ qubits \cite{GuhneToth2009}. Depth one means fully separable; depth two allows entanglement within pairs but does not establish GME for three or more parties.
For $J\subseteq\{1,\ldots,m\}$, let $M_J$ denote the corresponding marker subsystem and let $J^c=\{1,\ldots,m\}\setminus J$ be the complementary index set. For a balanced path input, define the normalized marker states conditioned on path $a$ and their reductions by
\begin{align}
 \rho_{M|a}&=2\langle a|\rho|a\rangle,\nonumber\\
 \rho_{M_J|a}&=\Tr_{M_{J^c}}\bigl(\rho_{M|a}\bigr),\nonumber\\
 D_J&=\tfrac12\|\rho_{M_J|0}-\rho_{M_J|1}\|_1.
 \label{eq:D}
\end{align}
Thus the partial trace removes precisely the markers outside $J$; $D_M$ denotes distinguishability with access to all markers. The trace norm is $\|R\|_1=\Tr\sqrt{R^\dagger R}$. With equal path priors, the optimal minimum-error guessing probability is $P_{\mathrm{guess}}^{\mathrm{opt}}(J)=(1+D_J)/2$ \cite{Helstrom1969}. Finding this optimum can require a collective measurement on $M_J$. Here $D_J$ quantifies minimum-error discrimination of the \emph{accessible} marker states; it is not the unambiguous-discrimination measure used in earlier multipath duality work \cite{SiddiquiQureshi2015,SiddiquiQureshi2021}. A purification-based measure can also differ from $D_J$ for mixed markers \cite{AbeEdamatsu2026}.

For a specified product positive-operator-valued measurement (POVM) $\{E_t\}$ on the markers, the unnormalized conditional path state is
\begin{equation}
 \tau_t=\Tr_M[(\id_A\otimes E_t)\rho]=q_t\rho_{A|t},
 \qquad q_t=\Tr\tau_t.
 \label{eq:tau}
\end{equation}
For $q_t>0$, $\rho_{A|t}$ is the normalized state conditioned on outcome $t$. These matrices specify the actual interference record. Passing the conditioned path through a balanced recombiner with phase shift $\varphi$, the probability of output $+$ is
\begin{equation}
 P(+|t,\varphi)=\tfrac12\left[1+2\operatorname{Re}\bigl(e^{i\varphi}\rho_{A|t}(0,1)\bigr)\right].
 \label{eq:conditionalfringe}
\end{equation}
The modulus of the off-diagonal element fixes the fringe visibility, while its phase fixes the fringe displacement; the corresponding joint probability is $P(+,t|\varphi)=q_tP(+|t,\varphi)$. Thus retaining the complex conditional coherence preserves more information than visibility alone. The conditional, unconditional, and average recovered visibilities are, respectively,
\begin{align}
 V_{A|t}&=2|\rho_{A|t}(0,1)|,\nonumber\\
 V_A&=2|\rho_A(0,1)|,\nonumber\\
 \Vrec&=\sum_tq_tV_{A|t}.
 \label{eq:visibilities}
\end{align}
For the balanced path populations, the conventional two-path predictability is $\mathcal P=|\rho_A(0,0)-\rho_A(1,1)|=0$. This does not force $D_J=0$; the marker states conditioned on the two paths may still be distinguishable. For a two-path state, $V_A$ equals its normalized $l_1$ coherence in the path basis, whereas $\Vrec$ refers to the conditioned eraser subensembles.
For a physical example, take a path qubit $A$ and two markers $B,C$ measured in $Z_B$ and $X_C$. In the three-qubit preparations studied in Sec.~III, with path coherence parameter $0\le\eta\le1$ and no white noise, each outcome $(b,x)$ has probability $1/4$ and leaves the path in the state
\begin{equation}
 q_{b,x}=\frac14,\qquad
 \rho_{A|b,x}=\frac12[\id_A+(-1)^b\eta X_A].
 \label{eq:threequbiteraser}
\end{equation}
The $b=0,1$ subensembles have equally visible fringes shifted by $\pi$; the $X_C$ outcome $x$ does not change them. As we show below, these fringes can arise from preparations with different entanglement depths. The measured data therefore need not identify the source.

Outcomes with $q_t=0$ contribute zero. We call $\mathcal D_E(\rho)=\{\tau_t(\rho)\}_t$ the \emph{eraser-tomography data}: they contain the probability of each marker outcome together with its unnormalized conditional path state. Thus they determine fringe amplitudes and phases \emph{and} path populations; a phase scan alone is insufficient for the latter. When optimal path distinguishabilities have been measured separately, we write $\mathcal D_{E,D}(\rho)=(\mathcal D_E(\rho),\{D_J(\rho)\}_{J\subseteq\{1,\ldots,m\}})$. Specifying $D_J$ does not amount to retaining every outcome of a particular discrimination measurement.

A \emph{compatible state} reproduces every specified statistic, including the balanced path probabilities. We write $\mathcal S(\mathcal D)$ for the states compatible with data $\mathcal D$ and ask for their least entanglement depth,
\begin{equation}
 k_{\min}(\mathcal D)=\min_{\rho\in\mathcal S(\mathcal D)}k(\rho),
 \label{eq:mindepthdef}
\end{equation}
where $k(\rho)$ is the depth of $\rho$. The information added at each stage is important. Eraser tomography fixes the conditional path states; separately obtained $D_J$ values add optimal path-guessing information; parity and full product-Pauli records constrain additional correlations. The tilted-axis test below uses $\mathcal D_E$ and one added complete product-outcome distribution, \emph{without} assuming independently measured $D_J$. A \emph{marker configuration} fixes the marker axes, whereas a \emph{joint local measurement setting} chooses one axis on each qubit. Complete eraser tomography generally needs separate $X_A$, $Y_A$, and $Z_A$ readouts with the same marker configuration. A parity expectation contains less information than its joint outcome distribution, and the recovered visibility $\Vrec$ cannot be substituted for unconditional $V_A$ in the standard duality relation.

For one orthonormal product-projective marker basis, the eraser-tomography data have a fully separable realization,
\begin{equation}
 \rho_{\rm cl}=\sum_t\tau_t\otimes|t\rangle\langle t|_M.
 \label{eq:recordseparable}
\end{equation}
This state is normalized and reproduces every $\tau_t$; it is obtained by local dephasing (a nonselective projective measurement) in the marker basis \cite{HorodeckiEB2003}. Thus $\mathcal D_E$ alone cannot certify entanglement. The construction does not generally preserve the additional optimal distinguishabilities $D_J$ and does not apply to an arbitrary nonprojective marker POVM.

\section{Entanglement from eraser data and path discrimination}
We now give a concrete ambiguity: one source entangles all qubits, while another contains entangled pairs only. They lead to the same eraser records even when the optimal path distinguishabilities are known.

With $|\Phi_\pm^{(m)}\rangle=(|0^m\rangle\pm|1^m\rangle)/\sqrt2$, put
\begin{equation}
 |G_m\rangle=\frac{|0\rangle_A|\Phi_+^{(m)}\rangle+|1\rangle_A|\Phi_-^{(m)}\rangle}{\sqrt2}.
 \label{eq:G}
\end{equation}
This is the standard $(m+1)$-qubit GHZ state after a Hadamard on $A$. One preparation starts from $|+\rangle_A|\Phi_+^{(m)}\rangle_M$ and applies $Z_1$ conditioned on the path value $a=1$. The two paths imprint opposite phases on a correlated marker state. No proper subset of markers reveals that phase, but their joint $X$ parity distinguishes the paths perfectly. Correlations in the initial marker state are essential here: for factorized path-conditioned marker states, zero distinguishability at every single marker would imply $D_M=0$. Let $\Lambda_\eta(R)=[(1+\eta)R+(1-\eta)Z_ARZ_A]/2$ describe path dephasing, with coherence-retention parameter $\eta$. The map $\Delta_{X_j}(R)=(R+X_jRX_j)/2$ describes a nonselective local projective measurement in the $X_j$ basis, equivalently local dephasing in that basis (\emph{pinching}). Define
\begin{align}
 \rho_{G,m}(\eta)&=\Lambda_\eta(|G_m\rangle\langle G_m|),\nonumber\\
 \rho_{F,m}(\eta)&=\bigl(\Delta_{X_m}\circ\cdots\circ\Delta_{X_2}\bigr)[\rho_{G,m}(\eta)].
 \label{eq:families}
\end{align}
We call these the GHZ ($G$) and flagged ($F$) preparations: $F$ mixes states of $AM_1$ with product $X$-basis flags on the other markers; see Supplemental Material, Sec.~\suppsec{2}. Both channels are completely positive and trace preserving for $0\le\eta\le1$. Global white-noise mixing is denoted by $\widetilde\rho_{j,m}(\eta,p)=(1-p)\rho_{j,m}(\eta)+p\id/2^{m+1}$, $j=G,F$, $0\le p\le1$.

\begin{theorem}[Depth ambiguity]\label{thm:ambiguity}
For the same marker configuration $Z_1X_2\cdots X_m$, the preparations $G$ and $F$ have identical $\mathcal D_{E,D}$. At $p=0$ their retained values are
\begin{align}
 &D_M=1,\quad D_J=0\ (J\subsetneq\{1,\ldots,m\}),\quad V_A=0,\nonumber\\
 &\tau_{b,\bm x}=2^{-m-1}[\id_A+\eta(-1)^bX_A],
 \label{eq:record}
\end{align}
where $b=0,1$ is the $Z_1$ outcome and $x_j=\pm1$ are X outcomes. For $\eta>0$, the $G$ family has depth $m+1$, whereas the $F$ family has depth exactly two. Equality of the retained data persists under the same white-noise mixing.
\end{theorem}

\begin{figure*}[tbp]
 \centering
 \includegraphics[width=0.94\linewidth]{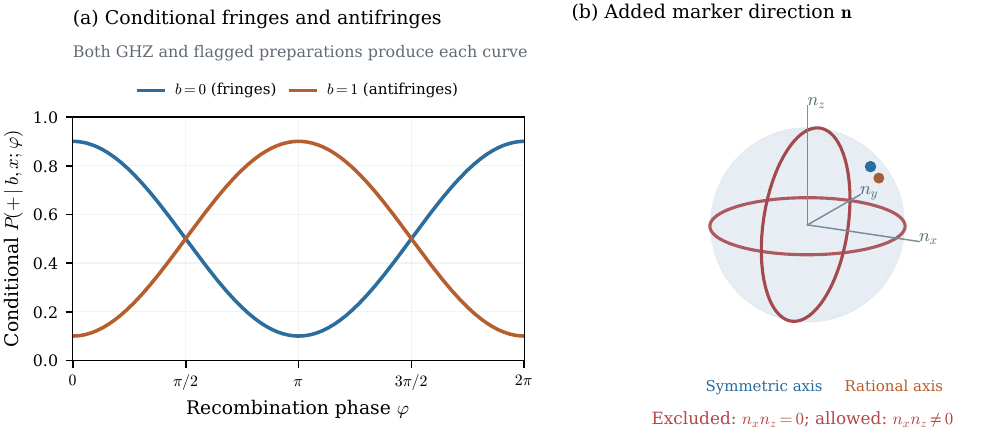}
 \caption{\textbf{Identical eraser fringes and measurement geometry.} (a) Conditional fringes for $\eta=0.8$ and $p=0$. Blue ($b=0$) and orange ($b=1$) denote the two $Z_B$ outcomes. Both GHZ and flagged preparations produce each curve, independently of the $X_C$ outcome, despite differing in entanglement depth. (b) For the pure three-qubit target, an additional product-projective configuration can certify genuine multipartite entanglement for marker-$C$ directions satisfying $n_xn_z\ne0$, with suitable measurement axes on $A$ and $B$. The circles show excluded directions; points mark the symmetric tilted and rational-witness axes.}
 \label{fig:geometry}
\end{figure*}

In the three-qubit example ($m=2$), the two $Z_B$ outcomes select a fringe and an antifringe, while sorting further by $X_C$ changes neither curve. They cancel in the unsorted signal, so $V_A=0$. Figure~\ref{fig:geometry}(a) shows the same conditional curves for the GHZ and flagged sources. Even access to all the listed $D_J$ values cannot distinguish their entanglement depths.

\emph{Proof sketch.} The $G$ conditional marker states are orthogonal, while all proper-subset reductions coincide. Nonselective local $X$-basis measurements \cite{HorodeckiEB2003} produce $\rho^F_{M|a}=2^{-m}[\id+(-1)^aX_1\cdots X_m]$ and preserve the eraser effects, proving the equality of the retained data. The flagged state F is a mixture of $AM_1$ pairs with orthogonal product X flags. Reading the flags recovers an entangled pair for $\eta>0$, proving that its depth is exactly two. By contrast, G has GHZ squared fidelity $(1+\eta)/2>1/2$, exceeding the maximum for any biseparable state. The decompositions and proofs are in the Supplemental Material, Secs.~\suppsec{1}--\suppsec{3}. \hfill$\square$

The maximal mean recovered visibility is $\Vrec=(1-p)\eta$ for both preparations, even with collective marker measurements, whereas $V_A=0$. In this construction, recovering the conditional fringes requires measuring only $M_1$, whereas optimal path discrimination requires access to all markers. At $\eta=1$, $p=0$, perfect recovered fringes and perfect distinguishability arise from different measurements on separate runs; this is consistent with ordinary duality. Their combination nevertheless constrains states separable across $A|M$: every balanced such state obeys
\begin{equation}
 \Vrec^{\,2}+D_M^2\le1.
 \label{eq:sepduality}
\end{equation}
The bound follows from the Bloch-ball constraint and convexity of the separable decomposition (Supplemental Material, Sec.~\suppsec{3}). Unlike ordinary path duality, Eq.~(\ref{eq:sepduality}) involves the \emph{recovered} visibility and is a separability condition. At $p=0$, the shared augmented data therefore certifies entanglement across $A|M$ for $\eta>0$.

White noise reduces both the optimal collective path distinguishability and recovered visibility: $D_M=1-p$ and $\Vrec=(1-p)\eta$. Set $s=1-p$ and $\nu=s\eta$. The separability inequality requires $s^2+\nu^2\le1$, and for these data it is also sufficient: the fully separable state
\begin{equation}
 \rho_{\mathrm{fs}}^{(E)}=2^{-m-1}
 [\id+\nu X_AZ_1+sZ_AZ_1\cdots Z_m]
 \label{eq:fsE}
\end{equation}
is fully separable whenever $s^2+\nu^2\le1$ and reproduces every eraser matrix and marker-subset distinguishability of the noisy GHZ family. Indeed, conditioning on the marker $Z$ basis gives product states whose path Bloch vectors have length $\sqrt{s^2+\nu^2}$. For $s^2+\nu^2>1$ no $A|M$-separable state reproduces the data, while the flagged state $F$, whose entanglement is confined to pairs, always does. Thus the \emph{minimum compatible entanglement depth} is exactly two for
\begin{equation}
 p<p_E(\eta):=1-\frac1{\sqrt{1+\eta^2}},
 \label{eq:pE}
\end{equation}
and exactly one at and above this boundary (Supplemental Material, Sec.~\suppsec{3}). Here $p_E$ refers to eraser data augmented by the supplied $D_J$, not to eraser data alone. The GHZ preparation remains genuinely multipartite entangled up to the established family threshold $p_{\mathrm{fam}}=\eta/(1+\eta-2^{-m})$ \cite{GuhneSeevinck2010,HanKye2021}. Hence for $\eta>0$ and $p_E\le p<p_{\mathrm{fam}}$, the prepared state is genuinely multipartite entangled, although a fully separable state reproduces all the retained data.

Recording an additional joint correlation can resolve more of the ambiguity. In particular, path--marker parity contains information absent from the list of optimal distinguishabilities $D_J$. The associated two-generator separability inequality is related to established stabilizer criteria \cite{TothGuhne2005,TothGuhnePRA2005}; here the boundary is made exact by constructing a matching fully separable preparation. If the full joint outcome distribution of the $Z_AX_1\cdots X_m$ measurement, or just its correlation, is \emph{also retained}, the exact compatible-depth threshold becomes
\begin{equation}
 p<p_{E,X}(\eta):=\frac{\eta}{1+\eta}.
 \label{eq:pEX}
\end{equation}
Below it every compatible state is entangled across $A|M$; at and above it an explicit fully separable state reproduces the data (Supplemental Material, Sec.~\suppsec{4}). For $m=2$ and $\eta=1$, using only $\mathcal D_{E,D}$ gives $p_E=1-1/\sqrt2$, whereas recording the parity correlation raises the threshold to $1/2$. At weak coherence, $p_E=\eta^2/2+O(\eta^4)$ while $p_{E,X}=\eta+O(\eta^2)$: retaining parity outcomes changes the leading sensitivity of path--marker entanglement inference from quadratic to linear in $\eta$. This does not by itself certify genuine multipartite entanglement, and it involves no postselection advantage.

\section{Genuine multipartite entanglement from additional measurements}
\subsection{Additional Pauli measurements}
We first examine one additional Pauli measurement using only the eraser data $\mathcal D_E$, without the independently supplied Helstrom distinguishabilities in $\mathcal D_{E,D}$. The subsequent two-Pauli converse uses the augmented data $\mathcal D_{E,D}$ and both full joint distributions. Throughout, additional settings are counted beyond the eraser tomography already specified.
\begin{theorem}[Limit of one additional Pauli measurement]\label{thm:pauli}
For every $m\ge2$ and every $\eta,p$, the original eraser data of $\widetilde\rho_{G,m}$ plus any one additional full product-Pauli configuration admits a biseparable explanation. For $p=0$, $\eta>0$, two additional configurations suffice.
\end{theorem}

\emph{Proof.} If the added configuration measures $Q\in\{X_2,Y_2,Z_2\}$ on $M_2$, perform a nonselective local projective measurement in the $Q$ basis, obtaining $\Delta_Q(\widetilde\rho_{G,m})$. The resulting state is separable across $M_2|\mathrm{rest}$ and preserves the added configuration. Dephasing in the $X$ basis fixes the baseline effects; dephasing in the $Y$ or $Z$ basis averages its equal $X$-outcome conditional matrices. For sufficiency, let
\begin{equation}
 \Ppar=\tfrac12(\id+Z_AX_1\cdots X_m),\quad
 \Pcorr=\prod_{j=1}^m\tfrac12(\id+X_AZ_j).
 \label{eq:projectors}
\end{equation}
The projector $\Ppar$ tests the joint path--marker parity, whereas $\Pcorr$ tests simultaneous satisfaction of the commuting path--marker correlations. Both pass probabilities come from complete product-Pauli outcome distributions. All biseparable states obey $\langle \Ppar+\Pcorr\rangle\le3/2$, while G gives $(3+\eta)/2$ at $p=0$. \hfill$\square$

To impose stronger constraints, supplement $\mathcal D_{E,D}$ with the full joint outcome distributions of \emph{two} specified Pauli configurations. The relevant operator $3\id-2(\Ppar+\Pcorr)$ is the established T\'oth--G\"uhne GHZ witness, up to a local Hadamard \cite{TothGuhne2005}. For the full joint outcome distributions of the specified noisy GHZ family, a biseparable explanation is excluded precisely when
\begin{equation}
 p<\frac{\eta}{2+\eta-2^{1-m}}.
 \label{eq:paulinoise}
\end{equation}
At equality and above, a GHZ-diagonal biseparable state reproduces the full joint outcome distributions of both product-Pauli measurements, every eraser matrix, and every $D_J$ (Supplemental Material, Sec.~\suppsec{4}). For this noisy GHZ family, the two complete Pauli distributions already determine the GME-certification boundary: adding the specified eraser matrices and independently supplied $D_J$ cannot improve that yes-or-no decision. The threshold is necessary and sufficient for these noisy GHZ statistics with the specified Pauli axes. If the parity-discrimination experiment was already recorded, its $Z_AX_1\cdots X_m$ outcomes determine $\Ppar$; only the $\Pcorr$ configuration is then additional (Supplemental Material, Sec.~\suppsec{4}).

The two pass probabilities $u=\Tr(\Ppar\rho)$ and $v=\Tr(\Pcorr\rho)$ also define a useful biseparable region. Its exact pentagon and the sharper attainable bound when the eraser data and $D_J$ are fixed are derived in Supplemental Material, Sec.~\suppsec{4}. Those scalar bounds do not replace the full-distribution converse stated above.

\subsection{Tilted-axis certification}
The restriction to Pauli axes matters. A single additional measurement along a \emph{tilted} marker direction can already rule out biseparable explanations of the pure three-qubit eraser. The reason is that the perfect $A$--$B$ correlation leaves only a two-dimensional joint subspace, behaving as a logical qubit. The new marker measurement probes this logical qubit in a way that the original eraser basis cannot. Geometrically, both the $X$ and $Z$ components of the added $C$ axis are needed; a $Y$ component is allowed.
\begin{theorem}[Tilted-axis certification criterion]\label{thm:tilt}
For the pure three-qubit target $|G_2\rangle$, keep the complete $Z_BX_C$ eraser data. Fix a unit Bloch direction $\bm n=(n_x,n_y,n_z)$ measured on $C$ in one added product-projective configuration. Some choice of the added axes on A and B certifies GME if and only if $n_xn_z\ne0$. All alternative explanations may be mixed.
\end{theorem}

\emph{Proof.} For necessity, perform a nonselective measurement on C along the added direction (local dephasing). This fixes all effects in the added configuration and makes the state separable across $C|AB$. The dual channel sends $X_C$ to $n_x(n_xX_C+n_yY_C+n_zZ_C)$. For the target's baseline, the conditional $X_C,Y_C$ correlations vanish, so the eraser data are unchanged when $n_xn_z=0$. The excluded directions form two great circles of the marker Bloch sphere, shown in Fig.~\ref{fig:geometry}(b).

Conversely, when $n_xn_z\ne0$, the eraser record contains the perfect correlation $\langle X_AZ_B\rangle=1$. This restricts $AB$ to an effective two-level system, or logical qubit. The added outcomes determine its states conditioned on the two $C'$ results. Positivity forces both states to be pure. A biseparable alternative must then predict a $Z_BX_C$ correlation proportional to $n_xn_z$, although the original eraser data fix that correlation to zero. The following calculation establishes the contradiction.

Let $n_{\perp}=\sqrt{n_x^2+n_y^2}>0$ be the transverse magnitude of $\bm n$, and write $\bm\sigma_C=(X_C,Y_C,Z_C)$. Choose
\begin{align}
 A'&=(X_A+Z_A)/\sqrt2,\qquad C'=\bm n\cdot\bm\sigma_C,\nonumber\\
 B'&=(n_xX_B-n_yY_B)/(\sqrt2\,n_{\perp})+Z_B/\sqrt2.
 \label{eq:idealaxes}
\end{align}
The perfect correlation $\langle X_AZ_B\rangle=1$ confines all compatible states to the $+1$ eigenspace, spanned by $|0_L\rangle=|+0\rangle_{AB}$ and $|1_L\rangle=|-1\rangle_{AB}$. These two states form one effective \emph{logical} qubit L. Within this support, a pure state product across $A|BC$ or $B|AC$ must have matching definite $X_A$ and $Z_B$ eigenvalues, and hence is fully product; states product across $C|AB$ are $C|L$ product. Therefore any biseparable compatible state is separable across $L|C$.

Next, the added joint outcome probabilities constrain two conditional logical-qubit observables. Within the logical subspace, $A'$ projects to $Z_L/\sqrt2$ and $A'B'$ to $(\id_L+E_L)/2$, where $E_L=(n_xX_L-n_yY_L)/n_{\perp}$ is a unit Bloch observable. The added measurement has two outcomes, $+$ and $-$, on C. Their probabilities and conditional logical-qubit expectations are
\begin{equation}
 q_{\pm}=\tfrac12,\qquad \langle Z_L\rangle_{\pm}=\pm n_z,\qquad
 \langle E_L\rangle_{\pm}=\pm n_{\perp}.
 \label{eq:logicaldata}
\end{equation}
The added measurement fixes two orthogonal Bloch components of each conditional logical state. Their squared lengths sum to $n_z^2+n_{\perp}^2=1$, the maximum allowed by positivity. Thus each conditional logical state is pure even when the original three-qubit preparation is not assumed pure. This use of the Bloch-ball boundary is related to all-versus-nothing steering arguments \cite{Chen2013}:
\begin{equation}
 P_{\pm}^L=\tfrac12[\id_L\pm(n_xX_L-n_yY_L+n_zZ_L)].
 \label{eq:logicalpure}
\end{equation}
The argument uses the complete, trusted eraser matrices, not conditional fringe visibility alone.

Finally, consider a putative biseparable explanation. The support restriction already makes it $L|C$ separable. In such a decomposition, each term contributing to a pure conditional state must have that logical pure factor. Since $P_+^L$ and $P_-^L$ are distinct, a term cannot contribute to both outcomes; its C state must be the corresponding eigenstate of $C'$. The unique separable explanation is consequently $\frac12(P_+^L\otimes\Pi_+^{C'}+P_-^L\otimes\Pi_-^{C'})$, where $\Pi_{\pm}^{C'}=(\id_C\pm C')/2$. It gives $\langle Z_LX_C\rangle=n_zn_x\ne0$, contradicting the baseline value $\langle Z_BX_C\rangle=0$. This proves sufficiency. \hfill$\square$

For $\bm n=(1,0,1)/\sqrt2$, all three added axes equal $(X+Z)/\sqrt2$. The eraser data alone fails to certify GME by Theorem~\ref{thm:ambiguity}, whereas this one additional product-projective setting succeeds. The distinguishing correlation $n_xn_z$ vanishes on approach to the excluded circles. Thus existence of a certifying setting for every nonexcluded direction is an exact-statistics statement; it does not imply a nonzero uniform finite-sample margin (Supplemental Material, Sec.~\suppsec{5}).

\paragraph{Minimum compatible entanglement for tilted measurements}
Certification is not the same as reconstructing a unique state. The added tilted data nevertheless permit exact lower bounds on the entanglement shared by \emph{all} compatible states. The conditioned logical states fix the diagonal blocks in the $C'$ outcome basis, leaving one complex coherence $c$ between the two outcome sectors. This coherence is \emph{not} the path coherence measured through fringe visibility. Positivity requires $|c|\le1/2$, and the eraser statistics impose $\kappa\operatorname{Re}c+n_y\operatorname{Im}c=\kappa/2$, where $\kappa=n_xn_z$. Thus the allowed values of $c$ form a line segment inside a disk; the explicit states and equality cases are given in Supplemental Material, Sec.~\suppsec{5}. For $n_xn_z\ne0$ and the added axes in Eq.~(\ref{eq:idealaxes}), the tight minima over compatible mixed preparations are
\begin{equation}
 \min C_{\rm GME}=\frac{|\kappa|}{\sqrt{\kappa^2+n_y^2}},\qquad
 \min f_G=\frac{\kappa^2}{\kappa^2+n_y^2}.
 \label{eq:exactquantification}
\end{equation}
Here $C_{\rm GME}$ is the convex-roof genuine-multipartite concurrence \cite{Ma2011} (zero on biseparable states and one for $|G_2\rangle$). It differs from the bipartite $I$-concurrence used in multipath triality \cite{Roy2022Triality}. The quantity $f_G=\langle G_2|\rho|G_2\rangle$ is the squared fidelity with the pure GHZ reference. The two minima are separate optimizations over the same compatible set. For $n_y\ne0$ they are generally attained by different states: the minimum concurrence occurs at the interior point $c_{\min}$, while the minimum squared fidelity occurs at an endpoint $c_{\rm end}$ (Supplemental Material, Sec.~\suppsec{5}). For $n_y=0$ and $\kappa\ne0$, the segment shrinks to a point: the target is uniquely identified. When $n_y\ne0$, distinct compatible states remain, even when the added A and B axes are varied. The corresponding identification statement uses the complete data specified above (Supplemental Material, Sec.~\suppsec{5}).

At the pure endpoint, $X_AZ_BX_C$ supplies the baseline constraints needed for this identification argument, so two full product contexts suffice in that ideal case. Two-context GHZ identification is established \cite{Yan2011}; the distinction here is the classification of completions of a \emph{fixed} eraser configuration. The corresponding finite-statistics limitations are treated below and in Supplemental Material, Sec.~\suppsec{5}.

\section{Noisy compatibility and entanglement certification}
\begin{figure*}[tbp]
 \centering
 \includegraphics[width=0.98\linewidth]{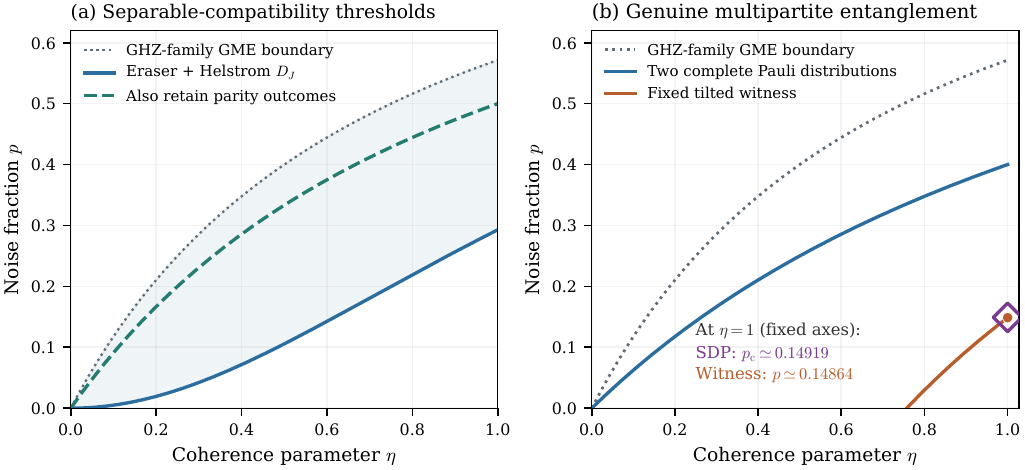}
 \caption{\textbf{How additional records change the noise threshold, for three qubits ($m=2$).} (a) Thresholds above which fully separable states reproduce the eraser data and independently supplied $D_J$, without ($p_E$) or with ($p_{E,X}$) recorded parity. The shaded region contains genuinely multipartite entangled GHZ-family targets whose supplied eraser/discrimination data admit fully separable explanations. (b) GHZ-family GME boundary, exact compatibility threshold for two complete Pauli distributions, and the sufficient tilted-witness threshold. At $\eta=1$, the orange circle denotes $p=0.148640865$ for the fixed witness; the purple diamond denotes the exact-SDP bracket $0.14919347\le p_{\mathrm c}\le0.14919348$ for complete eraser tomography plus one tilted distribution. The two markers nearly coincide at this scale. The single SDP point does not define a curve in $\eta$.}
 \label{fig:noise}
\end{figure*}
\subsection{Noise and excluded marker directions}
The pure-state axis classification can be extended to the noisy three-qubit family without selecting a particular witness. Keep the complete phase-resolved $Z_BX_C$ eraser data, and add one trusted product measurement
$A'=\bm a\cdot\bm\sigma_A$, $B'=\bm b\cdot\bm\sigma_B$, and $C'=\bm n\cdot\bm\sigma_C$. Let $\mathcal D_{\bm a,\bm b,\bm n}(\eta,p)$ contain the eraser matrices and all eight outcome probabilities of this added configuration. 
\begin{proposition}[All-noise marker-axis obstruction]\label{prop:noisyaxis}
For all $0\le p,\eta\le1$ and arbitrary added axes on $A$ and $B$, if $n_xn_z=0$ then $\mathcal D_{\bm a,\bm b,\bm n}(\eta,p)$ admits a state separable across $C|AB$. Consequently these directions cannot certify genuine tripartite entanglement, even with noiseless statistics. The same conclusion holds for every direction when $\eta=0$.
\end{proposition}
The first assertion follows by nonselectively measuring $C'$ and discarding its outcome. This local dephasing preserves the full added joint outcome distribution and, when $n_xn_z=0$, every conditional eraser matrix. At $\eta=0$ the noisy target is itself separable across $A|BC$. Conversely, for $\eta=1$ and $n_xn_z\ne0$, the axes constructed in Theorem~\ref{thm:tilt} give $p_{\rm c}>0$: the excluded pure-state data remain outside the closed biseparable-compatible set under sufficiently small white noise. The locally dephased alternative also gives an exact measurement-distance formula. Using the total-variation distance $\operatorname{TV}(P,Q)=\tfrac12\sum_x|P(x)-Q(x)|$, for the full $X_AZ_BX_C$ outcome distribution we find
\begin{equation}
 \operatorname{TV}(P_{\rho},P_{\omega})=\frac{(1-p)|n_xn_z|}{2},
 \label{eq:noisyTVmain}
\end{equation}
Here $\omega=\Delta_{C'}(\widetilde\rho_{G,2})$ is $C|AB$-separable. This distance concerns one explicitly constructed state, not the nearest biseparable state. For this specified comparator, statistical separation diminishes near the excluded directions (Supplemental Material, Sec.~\suppsec{6}); the expression does not determine an optimal sample cost.

\subsection{Exact noisy compatibility at fixed axes}
White noise weakens the correlations that separate the measured GHZ data from biseparable alternatives. We therefore seek the first noise fraction at which \emph{some} biseparable state reproduces the complete $Z_BX_C$ eraser tomography and all eight outcomes of one additional $A'B'C'$ measurement. The local axes are fixed and trusted. This test does not assume separately measured Helstrom distinguishabilities.

The main simplification is specific to these measured observables. In adapted local Pauli frames, their nonzero real moments are unchanged under partial transposition on any single-qubit cut. The remaining four baseline moments, containing $\widetilde Y_A$, vanish and are preserved by the corresponding real symmetrization. As proved in the Supplemental Material, every compatible biseparable state can therefore be represented by three unnormalized positive cut contributions $Q_A,Q_B,Q_C$ satisfying the constraints below. Conversely, these constraints construct a biseparable state. This makes the SDP \emph{exact}, rather than merely a relaxation.

More precisely, rotate the local frames so that the baseline and added axes lie in local $\widetilde X\widetilde Z$ planes. Let $O_i$ denote the independent traceless real measured observables, \emph{including zero-expectation constraints}, and set $d_i=\Tr[O_i\widetilde\rho_{G,2}(\eta,0)]$.

For $\bm d\ne0$, the largest compatible signal is
\begin{align}
 s_{\mathrm c}=\max_{s,Q_A,Q_B,Q_C}\quad&s,\nonumber\\
 Q_j&=Q_j^{\mathsf T}\succeq0,\quad Q_j^{\Gamma_j}=Q_j,\nonumber\\
 \Tr\sum_jQ_j&=1,\qquad
 \Tr\bigl(O_i\sum_jQ_j\bigr)=s d_i\quad(\forall i).
 \label{eq:exactNoisySDP}
\end{align}
Here $\Gamma_j$ denotes partial transposition on qubit $j$ in the adapted basis, and $s$ is not restricted to the physical interval during optimization. The \emph{exact} critical white-noise fraction is
\begin{equation}
 p_{\mathrm c}=\max\{0,1-s_{\mathrm c}\}.
 \label{eq:exactNoisyPC}
\end{equation}
For fixed axes, the recorded data admit a biseparable explanation exactly when $p\ge p_{\mathrm c}$; if $\bm d=0$, the maximally mixed state already supplies one and $p_{\mathrm c}=0$. The converse is constructive: each feasible $Q_j$ is separable across its cut because it is positive and invariant under partial transposition on that qubit \cite{Kraus2000}. A boundary state can be decomposed into at most 24 pure cut-product preparations, and mixing it with $\id/8$ reaches every larger physical noise fraction (Supplemental Material, Sec.~\suppsec{6}). The exactness refers to this \emph{fixed measurement dataset}; it does not imply optimality over all possible settings.

For a concrete comparison, fix the rational added axes
\begin{equation}
 T_A=\frac{3X_A+4Z_A}{5},\qquad
 T_j=\frac{4X_j+3Z_j}{5}\quad(j=B,C),
 \label{eq:tiltedaxes}
\end{equation}
and set $\eta=1$. Exact rational primal and dual certificates enclose the compatible-state threshold:
\begin{equation}
 0.14919347\le p_{\mathrm c}\le0.14919348.
 \label{eq:exactRationalBracket}
\end{equation}
The lower certificate excludes biseparable explanations below its bound, while an explicit biseparable preparation reproduces the data at the upper bound. These rational endpoints bracket the exact compatibility threshold; a numerical estimate is given in Supplemental Material, Sec.~\suppsec{6}.

\subsection{A fixed tilted witness}
For the same axes, a fixed witness provides a simpler sufficient test. The score $H_T$ combines eleven correlations from the baseline $X_AZ_BX_C$ and $Z_AZ_BX_C$ readouts with six from the added $T_AT_BT_C$ context. Its 17 rational coefficients and exact cutwise certificate are given in Supplemental Material, Sec.~\suppsec{7}. Evaluating this witness uses three full product contexts. Acquiring the \emph{complete} phase-resolved eraser data as well generally also requires $Y_AZ_BX_C$, making four contexts from scratch; independent $D_J$ values are a further resource.

For every biseparable three-qubit state, including mixed states, an exact rational certificate establishes
\begin{equation}
 \mathcal Q_T(\rho):=\Tr(H_T\rho)\le\frac{4257}{5000}
 \qquad (\rho\in\Bisep).
 \label{eq:Qbound}
\end{equation}
The corresponding entanglement witness is $W_T=(4257/5000)\id-H_T$; a negative expectation of $W_T$ certifies genuine tripartite entanglement. For the target family the score is
\begin{equation}
 \mathcal Q_T(\widetilde\rho_{G,2})=
 (1-p)\left(\frac{2423}{6250}+\frac{38273}{62500}\eta\right).
 \label{eq:Q}
\end{equation}
Consequently one added tilted configuration certifies genuine tripartite entanglement if
\begin{equation}
 (1-p)\left(\frac{2423}{6250}+\frac{38273}{62500}\eta\right)>\frac{4257}{5000}.
 \label{eq:Qnoise}
\end{equation}
At $\eta=1$, the rational bound guarantees detection for $p<18581/125006\simeq0.148640865$, slightly below the full-data compatibility boundary in Eq.~(\ref{eq:exactRationalBracket}). Figure~\ref{fig:noise}(b) labels both thresholds; their separation is too small to resolve at the plotted scale. Supplemental Material, Sec.~\suppsec{7}, gives the cutwise rational certificate. The witness uses all eraser outcomes without discarding subensembles.

\subsection{Finite-trial certification}
A negative expectation of $W_T$ can be certified from independent and identically distributed trials of a setting-independent source. The combined-score Hoeffding bound \cite{Hoeffding1963} gives the conservative criterion
\begin{equation}
 \widehat{\mathcal Q}_T-
 \sqrt{\frac{\ln(1/\alpha)}{2}\sum_{k=X,Z,T}\frac{R_k^2}{N_k}}
 -\sum_kR_ke_k>\frac{4257}{5000},
 \label{eq:pooledtest}
\end{equation}
for false-positive probability at most $\alpha$. The labels $X$, $Z$, and $T$ refer, respectively, to the product contexts $X_AZ_BX_C$, $Z_AZ_BX_C$, and $T_AT_BT_C$; the first two are readouts of the eraser tomography and the third is the added tilted setting. Here $N_k$ is the predetermined number of trials in context $k$, $R_k$ the range of its single-trial score, and $e_k$ an independently established bound on its readout-distribution error. The sufficient trial count for this fixed witness is larger than that of a conventional two-Pauli test under matched ideal assumptions; no experimental resource advantage is established. The full allocation, numerical comparison, and limitations associated with calibration, losses, and additional tomography are given in Supplemental Material, Sec.~\suppsec{8}, and Fig.~\ref{fig:supp:trials}.

\section{Discussion and conclusions}

The GHZ and flagged constructions show why quantum erasure and entanglement certification must be distinguished. Although the two preparations yield identical conditional fringes and optimal which-path distinguishabilities, their entanglement depths differ. Recording parity correlations narrows the set of compatible preparations, while retaining complete Pauli-outcome distributions restricts it further. The explicit matching states establish sharp limits on what each collection of measurements can certify.

For a fixed three-qubit eraser, the choice of an additional marker axis has a simple geometrical role. Directions with $n_xn_z=0$ leave biseparable explanations compatible with the measured data, whereas directions with $n_xn_z\ne0$ allow genuine multipartite entanglement to be certified for suitable path and first-marker settings. For specified tilted axes, we also determine the separate minima of compatible concurrence and squared GHZ fidelity. Entanglement can therefore be certified without uniquely identifying the preparation, although the certification margin may become small as the added axis approaches an excluded direction.

For the white-noise three-qubit family considered here, the exact semidefinite program determines the biseparable-compatibility boundary at fixed local measurement axes. The explicit tilted witness provides a simpler sufficient test, which need not attain that boundary. Both rely on recorded outcomes. The analysis assumes trusted measurements and allows initially correlated markers; the finite-trial guarantees additionally require independent trials and calibrated error bounds. Losses, drift, and incomplete path tomography remain outside the present analysis. Recovered interference alone therefore does not establish entanglement; the additional measured correlations determine which entanglement claims can be supported.

\section*{Data availability}
The numerical data and code supporting the findings of this study are available from the author upon reasonable request.
\bibliography{references}

% ---- Supplemental Material ----
\clearpage
\onecolumngrid
\clearpage
\setcounter{section}{0}
\setcounter{subsection}{0}
\setcounter{equation}{0}
\setcounter{figure}{0}
\setcounter{table}{0}
\setcounter{theorem}{0}
\setcounter{proposition}{0}
\renewcommand{\thesection}{S\arabic{section}}
\renewcommand{\theequation}{S\arabic{equation}}
\renewcommand{\thefigure}{S\arabic{figure}}
\renewcommand{\thetable}{S\arabic{table}}
\renewcommand{\theHsection}{supplement.\arabic{section}}
\renewcommand{\theHequation}{supplement.\arabic{equation}}
\renewcommand{\theHfigure}{supplement.\arabic{figure}}
\renewcommand{\theHtable}{supplement.\arabic{table}}
\providecommand{\theHtheorem}{}
\renewcommand{\theHtheorem}{supplement.\arabic{theorem}}
\providecommand{\theHproposition}{}
\renewcommand{\theHproposition}{supplement.\arabic{proposition}}
\providecommand{\theHlemma}{}
\renewcommand{\theHlemma}{supplement.\arabic{lemma}}
\input{supplement_body}

\end{document}

%% file: supplement_body.tex
\begin{center}
	{\Large\bfseries Supplemental Material}\\[5pt]
	{\large Entanglement inference from quantum erasure and path discrimination}\\[5pt]
	Mohd Asad Siddiqui\\[3pt]
	{\small Department of Physics, Government Girls P.G. College,}\\
	{\small Sawai Madhopur, Rajasthan 322001, India}\\[3pt]
	\href{mailto:asad@ctp-jamia.res.in}
	{\texttt{asad@ctp-jamia.res.in}}
\end{center}

\noindent\textbf{Measurement information.}
Throughout the supplement, we distinguish the
marker-conditioned path matrices $\tau_t$, which
determine conditional fringes and path populations,
from the independently determined optimal
which-path distinguishabilities $D_J$ and the
complete joint outcome distributions of additional
local measurements. These quantities impose
different constraints on compatible preparations.
Sections~\ref{sec:supp:1}--\ref{sec:supp:4}
establish the eraser-data constructions and
entanglement-depth thresholds.
Section~\ref{sec:supp:5} derives the tilted-axis
criterion and the minimum compatible entanglement
measures, while Sec.~\ref{sec:supp:6} establishes
the exact white-noise biseparable-compatibility
boundary for fixed measurement settings.
Sections~\ref{sec:supp:7}--\ref{sec:supp:8}
present the fixed-witness and finite-trial analyses.

\section{Conditional states and visibility in quantum erasure}\label{sec:supp:1}
A state is genuinely multipartite entangled (GME) if it is not biseparable. The set $\operatorname{Bisep}$ consists of all mixtures of pure states product across at least one bipartition. The order of tensor factors is $A,M_1,\ldots,M_m$; $X_j,Y_j,Z_j$ are local Pauli matrices with eigenvalues $\pm1$. All quantities are dimensionless. The trace norm is $\|R\|_1=\Tr\sqrt{R^\dagger R}$. The path priors are balanced, $\rho_A(0,0)=\rho_A(1,1)=1/2$. The normalized path-conditioned marker states are $\rho_{M|a}=2\langle a|\rho|a\rangle$, with reduced states $\rho_{M_J|a}=\Tr_{M_{J^c}}(\rho_{M|a})$, where $J\subseteq\{1,\ldots,m\}$ and $J^c=\{1,\ldots,m\}\setminus J$. The Helstrom distinguishability is $D_J=\|\rho_{M_J|0}-\rho_{M_J|1}\|_1/2$, and the optimal balanced-prior guessing probability is $(1+D_J)/2$ \cite{Helstrom1969}. This quantity may require collective measurements on $M_J$; its value does not specify the corresponding joint measurement outcomes.

For a marker positive-operator-valued measurement (POVM) effect $E_t$, the unnormalized path matrix is
\begin{equation}
 \tau_t(a,b)=\Tr[E_t\langle a|\rho|b\rangle].
 \label{eq:taudef}
\end{equation}
The partial trace equals that with $\sqrt{E_t}$ on both sides of $\rho$, so $\tau_t\ge0$. With the phase convention of the main text, a balanced recombiner gives
\begin{equation}
 P(+,t|\varphi)=\tfrac12[q_t+2\Re(e^{i\varphi}\tau_t(0,1))],\qquad q_t=\Tr\tau_t.
\end{equation}
Writing $\tau_t=q_t\rho_{A|t}$, the conditioned visibility is $V_{A|t}=2|\rho_{A|t}(0,1)|$ for $q_t>0$, and the unconditional visibility is $V_A=2|\rho_A(0,1)|$. The mean recovered visibility is $\Vrec=\sum_tq_tV_{A|t}$. If $q_t=0$, positivity forces $\tau_t=0$; the conditional state and fringe phase are undefined, and this outcome contributes zero to $\Vrec$. Complete $\tau_t$ requires the path populations as well as both fringe quadratures; phase scans alone need not determine every entry. As in the main text, $\mathcal D_E=\{\tau_t\}_t$ denotes the eraser-tomography data, whereas $\mathcal D_{E,D}=(\mathcal D_E,\{D_J\}_{J\subseteq\{1,\ldots,m\}})$ additionally specifies the optimized Helstrom values. The unconditional visibility follows from $\mathcal D_E$. As in the main text, $\mathcal S(\mathcal D)$ comprises the states reproducing the complete data specification $\mathcal D$, and $k_{\min}(\mathcal D)$ is their least entanglement depth. Both definitions concern trusted measurements and include the balanced path priors.

Write $R_{01}^{(j)}(1)=\langle0|\rho_{j,m}(1)|1\rangle$, $j=G,F$. Path dephasing and global white noise give
\begin{equation}
 \langle0|\widetilde\rho_{j,m}(\eta,p)|1\rangle=(1-p)\eta R_{01}^{(j)}(1).
\end{equation}
For any marker POVM,
\begin{equation}
 \Vrec=2(1-p)\eta\sum_t|\Tr[E_tR_{01}^{(j)}(1)]|\le(1-p)\eta.
 \label{eq:eropt}
\end{equation}
At $\eta=1,p=0$ the sum is at most 1/2 because each conditional visibility is at most one and all probabilities sum to one. The common eraser attains the bound. The optimization includes joint marker POVMs. This average recovery must be distinguished from a visibility conditioned on a selected successful outcome \cite{EnglertBergou2000}.

\subsection*{A fully separable explanation of a projective eraser measurement}
For an orthonormal product marker basis, put $\rho_{\rm cl}=\sum_t\tau_t\otimes|t\rangle\langle t|$. Each $\tau_t$ is positive, $\sum_t\Tr\tau_t=1$, and each marker projector is product. A qubit spectral decomposition of each $\tau_t$ proves full separability. Orthogonality gives $\Tr_M[(\id_A\otimes|t\rangle\langle t|)\rho_{\rm cl}]=\tau_t$. This construction, a nonselective local projective measurement (local dephasing, or pinching) in the eraser basis, does not generally preserve the $D_J$. It applies to one product-projective context, not to an arbitrary informationally complete marker POVM.

\section{Different entanglement structures with identical eraser measurements}\label{sec:supp:2}
The alternative reference form
\begin{equation}
 |G_m\rangle=\frac{|+\rangle_A|0^m\rangle+|-\rangle_A|1^m\rangle}{\sqrt2}
 \label{eq:Galt}
\end{equation}
is normalized and equals a local Hadamard on A of the standard GHZ state. The path-dephasing channel $\Lambda_\eta$ has Kraus operators $\sqrt{(1+\eta)/2}\,\id$ and $\sqrt{(1-\eta)/2}\,Z_A$, and is completely positive and trace preserving for $0\le\eta\le1$. The pinches $\Delta_{X_j}$ are local mixtures of two unitaries, commute on distinct sites, and are rank-one measure-and-prepare channels on their measured qubits \cite{HorodeckiEB2003}.

Here $G$ and $F$ denote the GHZ and flagged preparations. The marker states conditioned on the two paths in $G$ are orthogonal projectors onto $|\Phi_{\pm}^{(m)}\rangle$. For a marker subset $J$ with $1\le |J|<m$,
\begin{equation}
 \rho^G_{M_J|a}=\tfrac12\bigl[(|0\rangle\langle0|)^{\otimes |J|}+(|1\rangle\langle1|)^{\otimes |J|}\bigr].
\end{equation}
When no marker is accessible, both conditional reduced states are the scalar $1$ and provide no which-path information. Applying the marker pinches gives
\begin{equation}
 \rho^F_{M|a}=2^{-m}[\id+(-1)^aX_1\cdots X_m],\qquad
 \rho^F_{M_J|a}=\id_J/2^{|J|}\quad(J\subsetneq\{1,\ldots,m\}).
 \label{eq:paritysupport}
\end{equation}
The uniform states on the two full parity sectors are normalized, orthogonal, and have equal proper reductions. Hence in both families $D_M=1$ and every proper $D_J=0$. Measuring every marker in X and processing its parity attains $D_M$ locally for these families. This is not a general assertion of local attainability of Helstrom discrimination.

For X eigenvalues $x_j=\pm1$, define the parity flag $\pi(\bm x)=\prod_{j=2}^m x_j$. Direct projection of Eq.~(\ref{eq:Galt}) gives
\begin{equation}
 \bigotimes_{j=2}^m{}_X\langle x_j|G_m\rangle=2^{-(m-1)/2}|\chi_{\pi(\bm x)}\rangle,
 \qquad |\chi_s\rangle=(|+0\rangle+s|-1\rangle)_{AM_1}/\sqrt2.
\end{equation}
The nonselective $X$ readout gives the explicit normalized decomposition
\begin{equation}
 \rho_{F,m}(\eta)=2^{-(m-1)}\sum_{\bm x}
 \Lambda_\eta(|\chi_{\pi(\bm x)}\rangle\langle\chi_{\pi(\bm x)}|)
 \otimes|\bm x\rangle_X\langle\bm x|.
 \label{eq:SMflaggeddecomposition}
\end{equation}
A source can prepare F by classically sampling $\bm x$, preparing the associated two-qubit state on $AM_1$, and adding product $X$-basis flags. No GHZ source is required for this simulation.

Every common eraser projector
\begin{equation}
 E_{b,\bm x}=|b\rangle\langle b|_1\otimes\bigotimes_{j=2}^m|x_j\rangle_X\langle x_j|
\end{equation}
is fixed by all applied pinches. Self-adjointness of pinching in the trace pairing therefore preserves every conditional path matrix, not just its absolute visibility. Explicitly,
\begin{equation}
 \tau_{b,\bm x}=2^{-m-1}[\id+\eta(-1)^bX_A],\qquad q_{b,\bm x}=2^{-m}.
 \label{eq:tauexplicit}
\end{equation}
The unconditional path state is $\id_A/2$. Summing all X flags gives $\tau_b=[\id+\eta(-1)^bX_A]/4$. The conditional fringes can therefore be recovered by measuring $M_1$ alone. The distributed access structure in this construction concerns path discrimination.

The two Schmidt coefficients of $|G_m\rangle$ across every nontrivial cut are $1/\sqrt2$. Thus its squared overlap with any biseparable state is at most 1/2. Equivalently the Hermitian fidelity witness
\begin{equation}
 W_G=\id/2-|G_m\rangle\langle G_m|
 \label{eq:fidelitywitness}
\end{equation}
has nonnegative expectation on every biseparable state. Since
\begin{equation}
 f_G(\rho_{G,m}(\eta)):=\langle G_m|\rho_{G,m}(\eta)|G_m\rangle=(1+\eta)/2,
\end{equation}
G is genuinely $(m+1)$-partite entangled for $\eta>0$. We use the squared pure-state fidelity $f_G$ (the direct overlap with the GHZ reference), rather than its square root. At $\eta=0$,
\begin{equation}
 \rho_{G,m}(0)=\tfrac12\sum_{a=0}^1|a\rangle\langle a|_A\otimes|\Phi_{(-1)^a}^{(m)}\rangle\langle\Phi_{(-1)^a}^{(m)}|,
 \label{eq:eta0}
\end{equation}
which is separable across $A|M$.

White-noise mixing leaves the path priors balanced and gives $\widetilde\rho_{M|a}=(1-p)\rho_{M|a}+p\id_M/2^m$. Therefore $D_M=1-p$, all proper $D_J=0$, and Eq.~(\ref{eq:tauexplicit}) acquires the factor $(1-p)$ multiplying $\eta$. Both states coincide at $p=1$. The mixture with white noise preserves two-producibility of F.

\subsection*{Parity outcomes and conditional fringe measurements}
A parity-based comparator is useful for separating equality of scalar recovery quantities from equality of complete conditional matrices. Let $R=M_1\cdots M_{m-1}$, let $s(\bm k)=\bigoplus_{j=1}^{m-1}k_j$, and define, in the fixed order $A,M_1,\ldots,M_m$,
\begin{align}
 |\psi_{\bm k}\rangle&=\frac{|0\rangle_A|\bm k\rangle_R|s(\bm k)\rangle_{M_m}
 +|1\rangle_A|\bm k\rangle_R|s(\bm k)\oplus1\rangle_{M_m}}{\sqrt2},\\
 \rho_{{\rm par},m}(\eta)&=\Lambda_\eta\!\left[2^{-(m-1)}\sum_{\bm k}|\psi_{\bm k}\rangle\langle\psi_{\bm k}|\right].
 \label{eq:paritycomparator}
\end{align}
The sum is a normalized positive mixture. Each component contains a two-qubit state on $AM_m$ and computational product flags on R, so the state is two-producible. Conditional on path a, its marker distribution is uniform over computational strings of total parity a. Thus $D_M=1$, all proper $D_J=0$, and $V_A=0$. Measuring R in Z and $M_m$ in X attains mean recovered visibility $\eta$, which is optimal by the off-diagonal scaling argument of Sec.~\ref{sec:supp:1}. The comparator uses a different eraser measurement for general $m$.

For $m=2$, write $B=M_1$, $C=M_2$ and use the same $Z_BX_C$ eraser for G and the parity comparator. Let $b=0,1$ and $x=\pm1$ be the B and C outcomes. Both give four outcomes of probability 1/4 and conditional visibility $\eta$. Their signed quadratures are nevertheless
\begin{equation}
 \langle X_A\rangle^{G}_{b,x}=\eta(-1)^b,\qquad
 \langle X_A\rangle^{{\rm par}}_{b,x}=\eta x.
 \label{eq:parityquadratures}
\end{equation}
Equivalently their conditional matrices are $[\id+\eta(-1)^bX_A]/2$ and $[\id+\eta xX_A]/2$. They differ on two of the four outcomes whenever $\eta>0$. At $\eta=0$ the contrast vanishes and no conditional phase is assigned. The flagged F preparation in Sec.~\ref{sec:supp:2} also matches the signed quadratures of the prescribed eraser data.

The simpler $\langle Z_BZ_C\rangle_G=1$, $\langle Z_BZ_C\rangle_F=0$ example at $p=0$ distinguishes the two stipulated G,F density matrices for every $\eta$. Under common white noise these values become $1-p$ and 0. It does not exclude other biseparable states with the measured correlator. State identification and arbitrary-state GME certification are different tasks.

\section{Separability bounds and entanglement inferred from measurements}\label{sec:supp:3}
\subsection*{Path--marker separability bound}
For a balanced state separable across $A|M$, write $\rho=\sum_kw_k\rho_A^{(k)}\otimes\sigma_M^{(k)}$. If the path Bloch vectors are $(x_k,y_k,z_k)$, then
\begin{align}
 D_M&=\left\|\sum_kw_kz_k\sigma_M^{(k)}\right\|_1\le\sum_kw_k|z_k|,\\
 \Vrec&=2\sum_t\left|\sum_kw_k\rho_A^{(k)}(0,1)\Tr(E_t\sigma_M^{(k)})\right|
 \le\sum_kw_k\sqrt{x_k^2+y_k^2}.
\end{align}
Positivity and POVM completeness justify the last inequality. The Euclidean triangle inequality and the Bloch ball give
\begin{equation}
 \sqrt{D_M^2+\Vrec^{\,2}}\le\sum_kw_k\sqrt{z_k^2+x_k^2+y_k^2}\le1.
 \label{eq:sepbound}
\end{equation}
Thus the common data can certify $A|M$ entanglement despite failing to certify genuine multipartite entanglement. For the target information set this necessary condition is also \emph{exact}: the reverse implication is provided by an explicit fully separable simulator below. Equation~(\ref{eq:sepbound}) concerns recovered visibility, rather than the ordinary duality relation with unconditional $V_A$.

\begin{theorem}[Exact least depth compatible with the eraser information]\label{thm:SMexactdepth}
For $m\ge2$, set $s=1-p$, $\nu=s\eta$ and $d=2^{m+1}$. The noisy target's entire $\mathcal D_{E,D}$ admits a state separable across $A|M$ if and only if it admits a fully separable state, if and only if $s^2+\nu^2\le1$. The minimum compatible entanglement depth is one in this domain, including equality, and two otherwise.
\end{theorem}
\begin{proof}
The target has $D_M=s$ and $V_{\rm rec}=\nu$, so Eq.~(\ref{eq:sepbound}) proves necessity. For sufficiency consider
\begin{align}
 \rho_{\mathrm{fs}}^{(E)}&=d^{-1}[\id+\nu X_AZ_1+sZ_AZ_1\cdots Z_m]
 =2^{-m}\sum_{\bm k}\rho_A(\bm k)\otimes|\bm k\rangle\langle\bm k|,\nonumber\\
 \rho_A(\bm k)&=\tfrac12[\id+\nu(-1)^{k_1}X_A+s(-1)^{\sum_jk_j}Z_A].
 \label{eq:SMfsE}
\end{align}
Every path Bloch vector has length $\sqrt{s^2+\nu^2}\le1$, giving an explicit convex decomposition into fully product states. The full spectrum is $(1\pm\sqrt{s^2+\nu^2})/d$, each with multiplicity $2^m$. Applying the eraser $Z_1X_2\cdots X_m$ removes the parity term for $m\ge2$ and gives $\tau_{b,\bm x}=d^{-1}[\id+\nu(-1)^bX_A]$, exactly the target data. Direct path conditioning yields $\rho_{M|0}-\rho_{M|1}=2^{1-m}sZ_1\cdots Z_m$; its full Helstrom distance is $s$ and all proper-subset differences vanish. The off-diagonal path block is $\nu Z_1/d$, so the maximum mean recovered visibility also equals $\nu$. The noisy flagged preparation is two-producible and matches $\mathcal D_{E,D}$ for every parameter choice. Thus when $s^2+\nu^2>1$ the least depth is exactly two rather than merely at least two.
\end{proof}
The sharp path-cut inference threshold is
\begin{equation}
 p<p_E(\eta)=1-(1+\eta^2)^{-1/2}.
 \label{eq:SMpE}
\end{equation}
By contrast, the established GHZ-diagonal criterion gives the actual target-family GME boundary $p<p_{\rm fam}(\eta,m)=\eta/(1+\eta-2^{-m})$ [Eq.~(\ref{eq:exactgme})]. For every $\eta>0$, $p_E<p_{\rm fam}$. To see this, put $a=1-2^{-m}$; the needed positive inequality $a+\eta>a\sqrt{1+\eta^2}$ follows on squaring, with difference $2a\eta+(1-a^2)\eta^2>0$. Throughout $p_E\le p<p_{\rm fam}$ the actual target is GME but an entirely fully separable state reproduces the retained $\mathcal D_{E,D}$. For $m=2$ and $\eta=1$, the interval is $[1-1/\sqrt2,4/7)$. The statement concerns the \emph{least compatible depth}, not the true depth of a source known to produce $\widetilde\rho_{G,m}$.

The two-generator stabilizer full-separability inequality is established \cite{TothGuhne2005}. Directly measured probabilities from the eraser plus one additional Pauli configuration can therefore certify some entanglement. Let $K=Z_AX_1\cdots X_m$ and $L_1=X_AZ_1$. Every fully separable state obeys
\begin{equation}
 \langle K+L_1\rangle\le1.
\end{equation}
For a product state, bound the remaining X components by one and apply Cauchy--Schwarz to $(z_A,x_A)$ and $(x_1,z_1)$. Convexity proves the mixed-state statement. The target without white-noise mixing gives $1+\eta$. The Pauli obstruction of the main text therefore concerns GME, not all entanglement.

\subsection*{Negativity at perfect collective path distinguishability}
Use $\mathcal N_{A|M}=(\|\rho^{T_A}\|_1-1)/2$ \cite{VidalWerner2002}. In path blocks write $\rho=\left(\begin{smallmatrix} A&R\\R^\dagger&B\end{smallmatrix}\right)$, with $\Tr A=\Tr B=1/2$. The condition $D_M=\|A-B\|_1=1$ is equivalent to orthogonality of the positive supports of A and B. Positivity of $\rho$ confines R to a map from $\operatorname{supp}B$ to $\operatorname{supp}A$. Under partial transpose on the path qubit, A and B retain their positive eigenvalues, while R appears on the otherwise empty subspaces as a block with eigenvalues equal to plus and minus its singular values. Consequently,
\begin{equation}
 \mathcal N_{A|M}(\rho)=\|R\|_1,\qquad
 \Vrec=2\sum_t|\Tr(E_tR)|\le2\|R\|_1.
 \label{eq:negproof}
\end{equation}
For the last inequality, choose phases $z_t$, $|z_t|=1$, to align the scalar traces. The unital completely positive map $\operatorname{diag}(z_t)\mapsto\sum_tz_tE_t$ is an operator-norm contraction; trace-norm duality completes the bound. In G, $R=(\eta/2)|\Phi_+\rangle\langle\Phi_-|$, so $\|R\|_1=\eta/2$. In F the product X flags form orthogonal blocks with total trace norm $\eta/2$. Both therefore attain the lower bound for the complete specified data and every $D_J$. This is exact at $D_M=1$; finite-error use requires a separate robust bound.

\subsection*{Exact full-separability threshold of the flagged comparator}
In the product X-flag basis, $\widetilde\rho_{F,m}$ is a uniform direct sum of two-qubit states
\begin{equation}
 \beta_t=(1-p)\Lambda_\eta(|\chi_t\rangle\langle\chi_t|)+p\id_4/4.
\end{equation}
The pair state at $p=0$ is locally equivalent to a mixture of two Bell states with weights $(1\pm\eta)/2$. Its partial-transpose eigenvalues after white noise are $p/4+(1-p)/2$ (twice) and $p/4\pm(1-p)\eta/2$. Consequently,
\begin{equation}
 \widetilde\rho_{F,m}\text{ has depth exactly two}\quad\Longleftrightarrow\quad
 p<\frac{2\eta}{1+2\eta}.
 \label{eq:flagthreshold}
\end{equation}
Below this threshold, measuring the product flags leaves an entangled pair, so the full state is not fully separable. At and above it, every $2\times2$ block has positive partial transpose (PPT) and hence is separable; expanding these blocks with the product flags proves full separability. Thus the complement of Eq.~(\ref{eq:flagthreshold}), including equality, is its exact full-separability region. The minimum global partial-transpose eigenvalue across A is $p/2^{m+1}-(1-p)\eta/2^m$.

\subsection*{Pure-state boundary and exact GHZ-family noise classification}
For a globally pure balanced state, suppose $D_M>0$ and every proper $D_J=0$. In a hypothetical product cut, put A with marker subset $J\subsetneq\{1,\ldots,m\}$. The opposite factor is a path-independent pure state, giving $\rho_{M|a}=\rho_{M_J|a}\otimes|\xi\rangle\langle\xi|$ and hence $D_M=D_J$, a contradiction. Global purity is an additional assumption; no mixed-state version is inferred from this proof.

The noisy G state is diagonal in a locally rotated GHZ basis. With $n=m+1$, its spectrum is
\begin{equation}
 \lambda_\pm=(1-p)(1\pm\eta)/2+p/2^n,\qquad
 \lambda_0=p/2^n\quad\text{(multiplicity $2^n-2$)}.
 \label{eq:spec}
\end{equation}
The established GHZ-diagonal criterion is biseparability if and only if the largest basis weight is at most 1/2 \cite{GuhneSeevinck2010,HanKye2021}. Since $\lambda_+$ is largest, the exact GME boundary for this stipulated family is
\begin{equation}
 p<\frac{\eta}{1+\eta-2^{-m}}.
 \label{eq:exactgme}
\end{equation}
For completeness, the capped probability simplex is the convex hull of distributions with two entries equal to 1/2. An equal mixture of two distinct GHZ basis vectors is biseparable: vectors on the same computational-complement pair yield a classical product mixture; otherwise divide sites according to agreement/disagreement of their representative strings. The vectors become two orthogonal logical Bell states across that cut, whose equal mixture is separable. This reconstructs the known criterion. It is not a new classification or a full-data compatibility frontier. Equality in Eq.~(\ref{eq:exactgme}) is biseparable.

\section{Additional Pauli measurements and compatible preparations}\label{sec:supp:4}
The main article compares which-path discrimination data with the additional information retained from joint Pauli outcomes. This section first proves the obstruction for one extra Pauli setting, then establishes attainable bounds and an exact converse for two complete distributions.
\begin{lemma}[Commuting local effects cannot certify entanglement across that cut]
If all effects ever measured on one qubit commute, all recorded terminal local-product probabilities admit a state separable across that qubit versus the rest, when mixed alternatives are allowed.
\end{lemma}
\emph{Proof.} Pinch that qubit in a common eigenbasis. The dual channel fixes all its measured effects. For rank-one pinching the state becomes $\sum_kp_k|k\rangle\langle k|\otimes\rho_{\mathrm{rest}|k}$, preserving the probabilities and making the designated cut separable. Pairing with arbitrary observables on the remaining systems also proves equality of their unnormalized conditional matrices. If every effect is proportional to the identity, any basis suffices. The claim concerns terminal measurements, not coherent sequential reuse, purity promises, or unmeasured optimized functionals. \hfill$\square$

For the eraser baseline $Z_1X_2\cdots X_m$, the complete noisy conditional matrices are independent of every X outcome. If one additional full Pauli configuration uses Q on $M_2$, pinch in Q. Its effects are fixed. For Q=X the baseline is fixed; for Q=Y or Z the dual pinch sends each baseline X projector to $\id/2$. Averaging equal conditional matrices preserves every baseline matrix. Thus the two data sets admit a state separable across $M_2|\mathrm{rest}$ at every $\eta$,p. This simulation does not generally preserve the externally supplied optimal hierarchy when Q=Y or Z.

The two configurations for $\Ppar$ and $\Pcorr$ are respectively $Z_AX_1\cdots X_m$ and $X_AZ_1\cdots Z_m$. Their responses are
\begin{align}
 u_G=u_F&=1-p/2,\\
 v_G&=(1-p)(1+\eta)/2+p/2^m,\\
 v_F&=[1+(1-p)\eta]/2^m.
\end{align}
The known stabilizer witness gives $u+v\le3/2$ on every biseparable state \cite{TothGuhne2005,TothGuhnePRA2005}. The target excess is $[\eta-p(2+\eta-2^{1-m})]/2$. For $\eta>0$,$p=0$, the no-go plus this witness proves the exact additional-Pauli count of two.

The local protocol attaining $D_M$ conditions on the path Z outcome and reads all marker X outcomes. Its parity-guessing success probability equals $\langle \Ppar\rangle$. If these actual trials are part of the baseline, u is already measured: the flagged comparator defeats zero further configurations, while one new $\Pcorr$ configuration certifies the $p=0$ G family. A scalar promised optimal $D_M$ and retained raw parity outcomes are different information resources. The corresponding setting counts are as follows. Relative to the complete eraser data and already performed path tomography, the exact $p=0$ \emph{additional} Pauli cost is two configurations for every $m\ge2$ and $\eta>0$. For the pure three-qubit target, one arbitrary product-projective configuration suffices. If the raw $\Ppar$ parity-discrimination outcomes are also retained, the Pauli cost drops to one added configuration for the $p=0$ family; the tilted three-qubit protocol likewise uses one. These numbers count additional configurations, not individual path readouts, calibration, or samples, and they do not establish globally minimal total cost from scratch.

\subsection*{How retaining parity outcomes strengthens certification}
\begin{theorem}[Parity-augmented data threshold]\label{thm:SMparityrecord}
Supplement $\mathcal D_{E,D}$ by the joint outcomes of the $Z_AX_1\cdots X_m$ measurement, or just the correlation $\langle K\rangle=s$ where $K=Z_AX_1\cdots X_m$. The minimum compatible depth is two if $s+\nu>1$ and one if $s+\nu\le1$, equivalently the data certify path-cut entanglement exactly when $p<\eta/(1+\eta)$.
\end{theorem}
\begin{proof}
The eraser matrices fix $\langle L_1\rangle=\nu$, where $L_1=X_AZ_1$. For every product state across $A|M$,
\begin{equation}
 \langle K+L_1\rangle=z_A\langle X_1\cdots X_m\rangle+x_A\langle Z_1\rangle\le1.
\end{equation}
Indeed, $(x_A,z_A)$ lies in the unit disk; the marker operators $X_1\cdots X_m$ and $Z_1$ anticommute, so the squared sum of their expectation values is at most one. Convexity handles separable mixtures, giving $s+\nu\le1$. Conversely, whenever $s+\nu\le1$,
\begin{equation}
 \rho_{\mathrm{fs}}^{(E,X)}=d^{-1}[\id+sK+\nu L_1]
 =(1-s-\nu)\frac{\id}{d}+s\frac{\id+K}{d}+\nu\frac{\id+L_1}{d},
 \label{eq:SMfsEX}
\end{equation}
where $d=2^{m+1}$, is a fully separable state because each normalized product-Pauli projector is an equal mixture of product eigenstates. The parity term disappears under the eraser, so every $\tau_{b,\bm x}$ is reproduced; full and proper-subset Helstrom values are $s$ and zero, as in Theorem~\ref{thm:SMexactdepth}. The full parity outcome distribution is $P(a,\bm x)=d^{-1}[1+s(-1)^a\prod_jx_j]$ and is reproduced as well. Finally the locally pinched two-producible comparator F matches all the added measurement data for arbitrary $p,\eta$, fixing the least depth at two above the boundary.
\end{proof}
For $\eta>0$ the threshold $p_{E,X}=\eta/(1+\eta)$ exceeds $p_E=1-(1+\eta^2)^{-1/2}$. The difference is especially marked at weak coherence. For every fixed $m\ge2$, with $a_m=1-2^{-m}$ and $\eta\to0^+$,
\begin{align}
 p_E&=\frac{\eta^2}{2}+O(\eta^4),\nonumber\\
 p_{E,X}&=\eta-\eta^2+O(\eta^3),\nonumber\\
 p_{\rm fam}&=\frac{\eta}{a_m}-\frac{\eta^2}{a_m^2}+O(\eta^3),\nonumber\\
 p_{\rm fam}-p_E&=\frac{\eta}{a_m}+O(\eta^2).
 \label{eq:SMweakcoherence}
\end{align}
These expansions follow directly from Eqs.~(\ref{eq:SMpE}), (\ref{eq:exactgme}), and the parity threshold above. They compare the entanglement of the intended source family with the minimum entanglement forced by two different datasets; none is a universal noise threshold for arbitrary preparations. In particular, retaining parity outcomes restores a leading linear threshold for path--marker entanglement inference, not necessarily for genuine multipartite entanglement. It is essential not to replace raw discrimination data by the abstract Helstrom optimum $D_M$ in the resource accounting.

\subsection*{Two Pauli pass probabilities and their biseparable region}
The familiar stabilizer projectors give a useful starting point. We first characterize the two pass probabilities, then impose the remaining eraser and path-discrimination data to obtain matching compatible-state constructions.
Define $K=Z_AX_1\cdots X_m$, $L_j=X_AZ_j$, $\Ppar=(\id+K)/2$, and $\Pcorr=\prod_j(\id+L_j)/2$. The generators commute: K and $L_j$ have two local anticommutations. Their common +1 space is the GHZ ray, $\Ppar\Pcorr=|G_m\rangle\langle G_m|$. The ranks are $2^m$ and 2. For commuting projections $P+Q\le\id+PQ$, giving $\langle \Ppar+\Pcorr\rangle\le3/2$ on biseparable states. Equality is attained by $|+\rangle_A|0^m\rangle$. After a Hadamard on A, this is exactly the standard two-setting stabilizer witness \cite{TothGuhne2005,TothGuhnePRA2005}. Replacing adjacent Z-pair generators with generators centered on A does not change their common code projector.

\begin{figure}[htbp]
 \centering
 \includegraphics[width=0.61\textwidth]{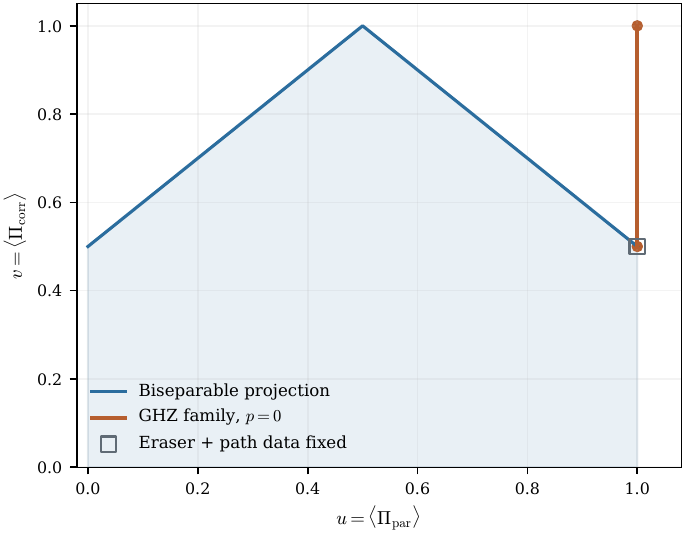}
 \caption{\textbf{Biseparable region of two Pauli pass probabilities.} The pentagon is the exact projection onto $(u,v)$, without fixing the complete eraser data. The orange segment gives the full $p=0$ GHZ target family, with endpoints $\eta=0,1$. The open square marks $v_{\max}=1/2$ at $u=1$ when the augmented eraser data are fixed.}
 \label{fig:pentagon}
\end{figure}

The rank-one projections $\Pcorr(\id\pm K)/2$ have biseparable overlap at most 1/2. Thus
\begin{equation}
 v+u\le3/2,\qquad v-u\le1/2.
 \label{eq:facets}
\end{equation}
The negative-K branch is $Z_1|G_m\rangle$, not the noise eigenvector $Z_A|G_m\rangle$. To prove exact attainment, choose a computational string $\bm w\notin\{0^m,1^m\}$ and its bitwise complement $\bm w^{\mathrm c}$, and $|\Psi_\pm\rangle=(|\bm w\rangle\pm|\bm w^{\mathrm c}\rangle)/\sqrt2$. The five $A|M$ product states
\begin{equation}
 |0,\Psi_-\rangle,\quad |0,\Phi_-\rangle,\quad |+,0^m\rangle,\quad |0,\Phi_+\rangle,\quad |0,\Psi_+\rangle
\end{equation}
give $(u,v)=(0,0),(0,1/2),(1/2,1),(1,1/2),(1,0)$. Their convex hull is precisely the claimed pentagon. These vertices can also be realized with balanced path populations: averaging any vertex state $\rho$ with $L_1\rho L_1$, where $L_1=X_AZ_1$, preserves both pass probabilities and makes $\langle Z_A\rangle=0$. Figure~\ref{fig:pentagon} illustrates this projection.
 Its interpretation as a two-dimensional projection is consistent with the capped GHZ-diagonal biseparable simplex \cite{HanKye2021}; it is not a new universal witness or the compatibility region for all joint outcome distributions.

\subsection*{Exact bound with the eraser data fixed}
\begin{theorem}[Maximum over biseparable states compatible with fixed noisy data]\label{thm:SMnoisyfrontier}
For $m\ge2$ and $p,\eta\in[0,1]$, fix every datum in $\mathcal D_{E,D}$ of the noisy target and also $u=\Tr(\Ppar\rho)=1-p/2$. Then
\begin{equation}
 \max_{\substack{\rho\in\Bisep,\ \mathcal D_{E,D}(\rho)=\mathcal D_{E,D}(\widetilde\rho_{G,m})\\ \Tr(\Ppar\rho)=1-p/2}}\Tr(\Pcorr\rho)
 =\frac{1+\min\{p,(1-p)\eta\}}2.
 \label{eq:SMnoisyfrontier}
\end{equation}
The optimum is attained by an explicit biseparable state, without restricting the search to GHZ-diagonal states.
\end{theorem}
\begin{proof}
Write $s=1-p$ and $\nu=s\eta$. The known universal facet $u+v\le3/2$ implies $v\le(1+p)/2$. Independently, the commuting projector inclusion $\Pcorr\le(\id+L_1)/2$ and the fixed eraser moment $\langle L_1\rangle=\nu$ give $v\le(1+\nu)/2$. These bounds apply to arbitrary mixed competitors.

For attainment, write $\Pi=|G_m\rangle\langle G_m|$ and introduce the four locally dephased states
$\omega_X=\Delta_{X_A}(\Pi)$, $\omega_Z=\Delta_{Z_A}(\Pi)$,
$\omega_Y=\Delta_{Y_A}(\Pi)$, and $\omega_m=\Delta_{X_m}(\Pi)$.
Each is biseparable because the nonselective projective measurement makes the pinched qubit a classical flag. Table~\ref{tab:dephasedalternatives} lists the values used in the attaining mixture.
\begin{center}
\begin{minipage}{0.90\textwidth}
\captionof{table}{Correlations and collective path distinguishability of four locally dephased alternatives used in the biseparable construction.}
\label{tab:dephasedalternatives}
\centering
\begin{tabular}{c@{\quad}cccc}
\toprule
State & $\langle L_1\rangle$ & $\langle K\rangle$ & $D_M$ & $\langle \Pcorr\rangle$ \\
\midrule
$\omega_X$ & 1&0&0&1\\
$\omega_Z$ & 0&1&1&$1/2$\\
$\omega_Y$ & 0&0&0&$1/2$\\
$\omega_m$ & 1&1&1&$1/2$\\
\bottomrule
\end{tabular}
\end{minipage}
\end{center}
Each has every proper-subset $D_J=0$; the eraser conditional matrices are $d^{-1}[\id+(-1)^b\langle L_1\rangle X_A]$ with $d=2^{m+1}$. Consider
\begin{equation}
 \rho_{\max}=\begin{cases}
 p\omega_X+(1-\nu)\omega_Z+(\nu-p)\omega_m,&p\le\nu,\\
 \nu \omega_X+(1-p)\omega_Z+(p-\nu)\omega_Y,&p\ge\nu.
 \end{cases}
 \label{eq:SMattainingmix}
\end{equation}
Every coefficient is nonnegative and the coefficients sum to one; both branches reproduce all $\tau_{b,\bm x}$, have $\langle K\rangle=s$, give $u=(1+s)/2=1-p/2$, and attain the stated $v$. The upper bound $D_M\le s$ follows from trace-norm convexity because the total weight of the $\omega_Z,\omega_m$ components equals $s$. The parity observable simultaneously gives $D_M\ge|\langle K\rangle|=s$. Thus $D_M=s$ exactly, and all proper $D_J=0$. In particular, the nonlinear optimized constraints have been checked rather than replaced by a linear moment.
\end{proof}
At $p=0$ the result is $v_{\max}=1/2$, recovering the previously derived noiseless slice. At $p=1$ the compatible-state optimum is also $1/2$, although the target maximally mixed state has $v=2^{-m}$. The two attaining branches meet at $p=\eta/(1+\eta)$. The theorem fixes $u=1-p/2$ and the specified eraser data.

\subsection*{Complete joint outcome distributions: the exact converse}
\begin{theorem}[Exact criterion from two complete Pauli outcome distributions]\label{thm:SMfullhist}
Retain the target's entire $\mathcal D_{E,D}$ and the full joint outcome distributions of the product-Pauli measurements $Z_AX_1\cdots X_m$ and $X_AZ_1\cdots Z_m$. A biseparable state matches all these data if and only if
\begin{equation}
 p\ge\frac{\eta}{2+\eta-2^{1-m}}.
 \label{eq:SMPauliExact}
\end{equation}
Thus the known T\'oth--G\"uhne witness is necessary and sufficient for excluding biseparable explanations \emph{on this specified target-data family}; it is not a universal new witness.
\end{theorem}
\begin{proof}
If Eq.~(\ref{eq:SMPauliExact}) is violated, the established facet $u+v\le3/2$ \cite{TothGuhne2005} rules out every biseparable model. To prove the converse while retaining the \emph{full} outcome distributions and all Helstrom values, use the locally rotated GHZ basis
\begin{equation}
 |G_{\pm,\bm w}\rangle=2^{-1/2}(|+\rangle_A|\bm w\rangle\pm|-\rangle_A|\bm w^{\mathrm c}\rangle),\qquad
 \Pi_{\pm,\bm w}=|G_{\pm,\bm w}\rangle\langle G_{\pm,\bm w}|,
\end{equation}
where $\bm w\in\{0,1\}^m$ and $\bm w^{\mathrm c}$ is its bitwise complement. The first product observable $K$ has eigenvalue $\pm1$ on these vectors; $L_j$ has eigenvalue $(-1)^{w_j}$. Choose
\begin{align}
 q_{\bm w}&=\frac{p}{2^m}+\frac{s}{2}[(1+\eta)\delta_{\bm w,0^m}+(1-\eta)\delta_{\bm w,1^m}],\\
 \ell_{\bm w}&=\max\{0,q_{\bm w}-1/2\},\qquad B=\sum_{\bm w}\ell_{\bm w}=\max\{0,q_{0^m}-1/2\},\\
 \alpha&=\frac{p/2-B}{1/2-B},\quad b_{\bm w}=\ell_{\bm w}+\alpha(q_{\bm w}/2-\ell_{\bm w}),\quad a_{\bm w}=q_{\bm w}-b_{\bm w},
 \label{eq:SMdiagweights}
\end{align}
where $s=1-p$. Under Eq.~(\ref{eq:SMPauliExact}), $B\le p/2$; the denominator is positive throughout the allowed domain. Thus $0\le\alpha\le1$, $0\le b_{\bm w}\le a_{\bm w}\le1/2$, $\sum b_{\bm w}=p/2$, and $\sum(a_{\bm w}-b_{\bm w})=s$. The normalized GHZ-diagonal state
\begin{equation}
 \rho_{\mathrm{match}}=\sum_{\bm w}(a_{\bm w}\Pi_{+,\bm w}+b_{\bm w}\Pi_{-,\bm w})
 \label{eq:SMdiagmatch}
\end{equation}
is biseparable by the established GHZ-diagonal criterion that its maximum basis probability does not exceed one half \cite{HanKye2021}. This criterion is stated and reconstructed in Sec.~\ref{sec:supp:3}; no novelty is claimed for it.

The full $X_AZ_1\cdots Z_m$ outcome distribution depends only on $q_{\bm w}=a_{\bm w}+b_{\bm w}$, which matches the target exactly. The full $Z_AX_1\cdots X_m$ outcome distribution depends only on $\sum_{\bm w}(a_{\bm w}-b_{\bm w})=s$ and is $P(a,\bm x)=d^{-1}[1+s(-1)^a\prod_jx_j]$. The eraser conditional matrices depend only on $\sum_{\bm w}(-1)^{w_1}q_{\bm w}=s\eta$, so every $\tau_{b,\bm x}$ matches the target. Finally, let $r_{\bm w}=a_{\bm w}-b_{\bm w}\ge0$. On each unordered complementary-string pair the Hermitian marker difference $\rho_{M|0}-\rho_{M|1}$ has eigenvalues $\pm(r_{\bm w}+r_{\bm w^{\mathrm c}})$, giving
\begin{equation}
 D_M=\sum_{\bm w}r_{\bm w}=s,
\end{equation}
while tracing out any marker kills each complementary-string coherence, so every proper $D_J=0$. This proves compatibility for the complete data sets, including the threshold equality.
\end{proof}
The converse applies to the stated target-data family with trusted measurements; Sec.~\ref{sec:supp:8} treats finite counts.

\begin{figure}[htbp]
 \centering
 \includegraphics[width=0.93\textwidth]{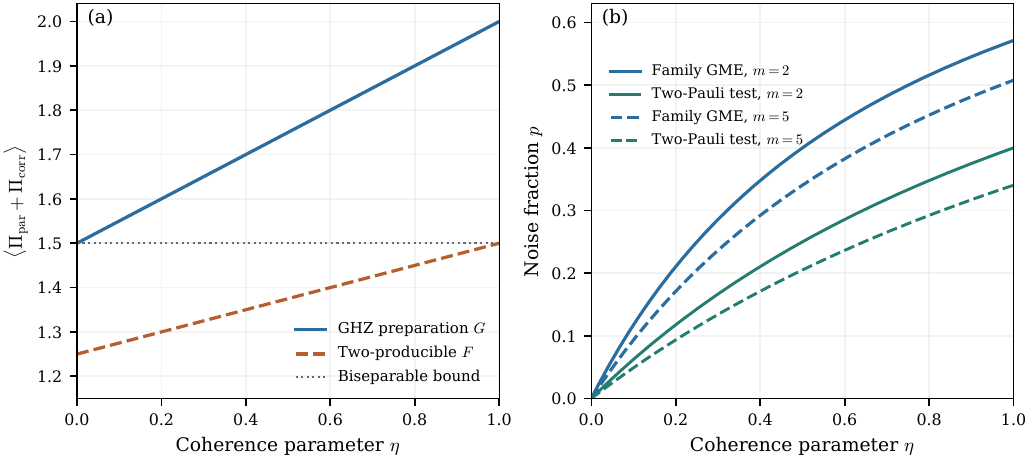}
 \caption{(a) Standard two-setting witness response at $m=2$,$p=0$ for G and the two-producible F comparator, relative to the tight biseparable bound. (b) Exact stipulated-family GME boundaries and standard two-setting detection thresholds for $m=2$ and $m=5$. The latter are witness-specific thresholds; the former follow from known GHZ-diagonal classification. The curves follow from the stated analytical thresholds.}
 \label{fig:certificationcomparison}
\end{figure}

Figure~\ref{fig:certificationcomparison} compares the standard witness response and its marker-number dependence.

\section{Tilted measurements and compatible entanglement}\label{sec:supp:5}
The main article gives a necessary-and-sufficient condition on the added marker direction. We prove the obstruction by local dephasing and the converse using the logical-qubit support fixed by the eraser data. We then describe all compatible states and the least entanglement they can contain.
Here $m=2$, $B=M_1$, $C=M_2$, and the target is $\rho=|G_2\rangle\langle G_2|$. Alternatives are arbitrary three-qubit mixed states.

\subsection*{Why certain marker directions cannot certify entanglement}
In every baseline test obtained by fixing B's Z outcome and any path observable, the correlations with local $X_C$ and $Y_C$ vanish; the only possible $Z_C$ component survives through the computational marker correlation. Since
\begin{equation}
 \Delta_{\bm n}^\dagger(X_C)=n_x(n_xX_C+n_yY_C+n_zZ_C),
\end{equation}
the pinched state preserves the baseline when $n_xn_z=0$. It preserves the added context because that context measures $\bm n\cdot\bm\sigma$ on C, and it is separable across $C|AB$. This proof permits arbitrary added A,B bases and also works for the noisy/dephased target. It proves the necessity direction of the main theorem.

\subsection*{Certification using the other marker directions}
Suppose $n_xn_z\ne0$, and put $n_{\perp}=\sqrt{n_x^2+n_y^2}>0$. One suitable choice is
\begin{equation}
 A'=(X_A+Z_A)/\sqrt2,\qquad
 B'=\frac{n_xX_B-n_yY_B}{\sqrt2\,n_{\perp}}+\frac{Z_B}{\sqrt2},\qquad
 C'=\bm n\cdot\bm\sigma_C.
 \label{eq:genericaxes}
\end{equation}
All are unit spin observables. The baseline fixes $\langle X_AZ_B\rangle=1$. Positivity therefore confines every compatible state to the support of $\Pi_L=(\id+X_AZ_B)/2$. Use the AB logical basis
\begin{equation}
 |0_L\rangle=|+0\rangle_{AB},\qquad |1_L\rangle=|-1\rangle_{AB}.
 \label{eq:logicalbasis}
\end{equation}
Within this support, $|G_2\rangle=(|0_L0\rangle+|1_L1\rangle)/\sqrt2$ and the baseline also fixes $\langle Z_LX_C\rangle=\langle Z_BX_C\rangle=0$.

Every pure-state component of a compatible mixture must be supported in the $+1$ eigenspace of $X_AZ_B$. If a component is product across A versus BC, $\langle X_A\rangle\langle Z_B\rangle=1$ forces A to be an X eigenstate and B to be the matching Z eigenstate. The pure component is fully product. The same argument holds across B versus AC. A component product across C versus AB is already $C|L$ product. Therefore every biseparable state in this support is separable across $L|C$.

Projected into the code, the measurable operators are
\begin{equation}
 A'\mapsto Z_L/\sqrt2,\qquad B'\mapsto Z_L/\sqrt2,\qquad
 A'B'\mapsto(\id_L+E_L)/2,\quad E_L=(n_xX_L-n_yY_L)/n_{\perp}.
 \label{eq:logicalmeasurements}
\end{equation}
The added full joint outcome distribution fixes these marginals for the two $C'$ outcomes. The target has $q_{\pm}=1/2$ and
\begin{equation}
 \langle Z_L\rangle_{\pm}=\pm n_z,\qquad \langle E_L\rangle_{\pm}=\pm n_{\perp}.
\end{equation}
As $n_z^2+n_{\perp}^2=1$ and the two directions are orthogonal, positivity forces the pure conditional logical density
\begin{equation}
 P_{\pm}^L=\tfrac12[\id_L\pm(n_xX_L-n_yY_L+n_zZ_L)].
 \label{eq:pureassemblage}
\end{equation}
No unmeasured orthogonal Bloch component can remain. The conditional-purity mechanism is related to all-versus-nothing steering arguments \cite{Chen2013}; here the eraser support and baseline moment determine its implications. This follows from the measured conditional states and does not assume that a competing global state is pure.

For a separable $L|C$ decomposition, every term contributing to a pure conditional state must have that logical pure factor. A term cannot contribute to both $C'$ outcomes because $P_+^L\ne P_-^L$. Its C factor must consequently have a deterministic $C'$ outcome. The unique separable explanation of these conditional states is
\begin{equation}
 \omega_{LC}=\tfrac12(P_+^L\otimes\Pi_+^{C'}+P_-^L\otimes\Pi_-^{C'}),\quad \Pi_{\pm}^{C'}=\tfrac12(\id_C\pm C').
 \label{eq:uniquesep}
\end{equation}
It predicts $\langle Z_LX_C\rangle=n_zn_x\ne0$, contradicting the baseline. This proves sufficiency and completes the exact iff geometry. Taking $\bm n=(1,0,1)/\sqrt2$ makes all three added observables $(X+Z)/\sqrt2$. The eraser-only data admit the flagged explanation, so one is also the exact additional arbitrary-projective count at this endpoint.

\subsection*{All compatible states and conditions for unique identification}
Let $U=(|n_+\rangle,|n_-\rangle)$ diagonalize $\bm n\cdot\bm\sigma$ and define $|n_s^{\mathrm{conj}}\rangle$ as the componentwise complex conjugate of $|n_s\rangle$ in the logical computational basis. The Bell identity
\begin{equation}
 |G_2\rangle=\frac{|f_+\rangle+|f_-\rangle}{\sqrt2},\qquad
 |f_s\rangle=|n_s^{\mathrm{conj}}\rangle_L|n_s\rangle_C
 \label{eq:bellrotatedbasis}
\end{equation}
is independent of eigenvector phases. The pure conditional logical states in Eq.~(\ref{eq:pureassemblage}) fix both diagonal blocks of a compatible state in the C measurement basis. Positivity then confines each off-diagonal block to the corresponding rank-one supports. Every compatible state must consequently have the form
\begin{equation}
 \rho(c)=\frac{|f_+\rangle\langle f_+|+|f_-\rangle\langle f_-|}{2}
 +c|f_+\rangle\langle f_-|+\overline{c}|f_-\rangle\langle f_+|,
 \qquad |c|\le\frac12.
 \label{eq:allcompatible}
\end{equation}
Here $c$ is coherence between the two eigenvalue sectors of the added $C'$ measurement, not the off-diagonal path element extracted from an eraser fringe. The reduced states on L and C are maximally mixed. Inside the logical support the baseline adds only $\langle Z_LX_C\rangle=0$. Since
\begin{equation}
 \langle f_-|Z_LX_C|f_+\rangle=-n_xn_z+i n_y,
\end{equation}
this condition is exactly
\begin{equation}
 \kappa\operatorname{Re}c+n_y\operatorname{Im}c=\kappa/2,\qquad\kappa=n_xn_z.
 \label{eq:chordconstraint}
\end{equation}
Conversely, every density operator satisfying these equations reproduces the complete collection of conditional logical states and the entire baseline. This proves necessity and sufficiency for the complete compatible set for the prescribed axes, without a purity promise on alternatives.

Partial transpose across C has its only negative eigenvalue $-|c|$. Thus the least compatible negativity across $C|AB$ is $|\kappa|/[2\sqrt{\kappa^2+n_y^2}]$, attained at
\begin{equation}
 c_{\min}=\frac{\kappa(\kappa+i n_y)}{2(\kappa^2+n_y^2)}.
 \label{eq:chordminimizer}
\end{equation}
The target overlap is $f_G=1/2+\operatorname{Re}c$. The other endpoint of the line segment,
\begin{equation}
 c_{\rm end}=\frac{\kappa^2-n_y^2}{2(\kappa^2+n_y^2)}
            +i\frac{\kappa n_y}{\kappa^2+n_y^2},
\end{equation}
attains the least compatible overlap $\kappa^2/(\kappa^2+n_y^2)$.

For genuine multipartite concurrence use the convention \cite{Ma2011}
\begin{equation}
 C_{\rm GME}(|\psi\rangle)=\min_{j=A,B,C}\sqrt{2[1-\Tr(\rho_j^2)]},
 \quad C_{\rm GME}(\rho)=\inf_{\{p_i,\psi_i\}}\sum_ip_iC_{\rm GME}(|\psi_i\rangle).
 \label{eq:gmeconcurrencedef}
\end{equation}
Any pure vector $a|f_+\rangle+b|f_-\rangle$ has $C_{C|AB}=2|ab|$ and $C_{A|BC}=C_{B|AC}=\sqrt{1-n_z^2(|a|^2-|b|^2)^2}\ge2|ab|$. Thus its GME concurrence is $2|ab|$. Every ensemble of $\rho(c)$ stays in its support and has average concurrence at least $2|c|$ by the triangle inequality. Equality is attained by equal mixing of two pure states with weights $t,1-t$ exchanged, a common relative phase, and $t=[1+\sqrt{1-4|c|^2}]/2$. Therefore
\begin{equation}
 C_{\rm GME}(\rho(c))=2|c|,\qquad
 \min_{\rho\ \mathrm{compatible}}C_{\rm GME}(\rho)=\frac{|\kappa|}{\sqrt{\kappa^2+n_y^2}}.
 \label{eq:tightgmeconcurrence}
\end{equation}
For $n_y=0$, $\kappa\ne0$, the allowed line touches the disk at one point and forces $c=1/2$: the target is uniquely identified. For $n_y\ne0$, $c_{\rm end}\ne1/2$ supplies a distinct compatible state for \emph{any} A,B readouts, since it reproduces the complete logical assemblage for this C measurement. On $\kappa=0$, pinching already gives a biseparable alternative. Hence some choice of A,B axes uniquely identifies the target if and only if $n_y=0$, $n_xn_z\ne0$. This fixed-baseline classification is distinct from established unrestricted two-observable GHZ identification \cite{Yan2011}.

\begin{figure}[htbp]
 \centering
 \includegraphics[width=0.85\textwidth]{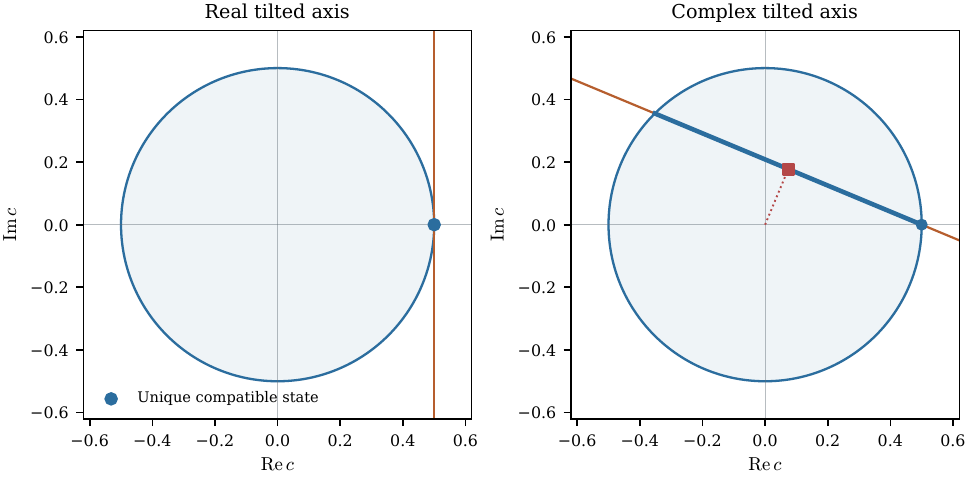}
 \caption{Exact compatible states for the axes of the tilted-axis theorem. The disk $|c|\le1/2$ imposes positivity; the orange line $\kappa\operatorname{Re}c+n_y\operatorname{Im}c=\kappa/2$ imposes the remaining eraser constraint. For a real tilted axis (left), the blue point marks the uniquely identified GHZ state. For a complex tilted axis (right), the thick blue line segment shows all compatible states and the red square marks the point of minimum genuine multipartite concurrence. No noise or global-purity promise is assumed for the alternatives.}
 \label{fig:compatiblechord}
\end{figure}

These are exact-data results. Along the real great circle the target remains unique arbitrarily close to a forbidden axis, while the statistical lower bound diverges. Exact identifiability therefore supplies no uniform robustness guarantee. Figure~\ref{fig:compatiblechord} illustrates the compatible-state geometry.

\section{Noisy compatibility and its exact semidefinite characterization}\label{sec:supp:6}
White noise can make a biseparable preparation compatible with the observed correlations before the intended GHZ family becomes biseparable. To locate that boundary, we retain complete phase-resolved $Z_BX_C$ eraser tomography and one additional $A'B'C'$ joint distribution, for arbitrary fixed local axes. The independently optimized Helstrom quantities in $\mathcal D_{E,D}$ are not assumed. We show that the resulting semidefinite program is exact for these measurements and provide rational certificates for one choice of axes. The later statistical comparison concerns specified locally dephased alternatives, not the complete biseparable set.

\subsection*{Noisy measurements and biseparable compatibility}

Let $s=1-p$ and write $A'=\bm a\cdot\bm\sigma_A$, $B'=\bm b\cdot\bm\sigma_B$, $C'=\bm n\cdot\bm\sigma_C$, with unit Bloch vectors. In the Pauli basis, the noisy target has the exact expansion
\begin{align}
8\widetilde\rho_{G,2}(\eta,p)={}&\id+s\bigl[Z_BZ_C+Z_AX_BX_C-Z_AY_BY_C\nonumber\\
 &+\eta(X_AZ_B+X_AZ_C+Y_AX_BY_C+Y_AY_BX_C)\bigr].
 \label{eq:allaxespauli}
\end{align}
Of the traceless moments available in the baseline eraser configuration with path tomography, only $\langle X_AZ_B\rangle=s\eta$ is nonzero for the target. The seven nontrivial marginals and correlations of $A'B'C'$ have vanishing single-qubit expectations, while
\begin{align}
 \langle A'B'\rangle&=s\eta a_xb_z, &
 \langle A'C'\rangle&=s\eta a_xn_z,\nonumber\\
 \langle B'C'\rangle&=s b_zn_z,&
 \langle A'B'C'\rangle&=s\bigl[\eta a_y(b_xn_y+b_yn_x)\nonumber\\
 & & &\hspace{11mm}+a_z(b_xn_x-b_yn_y)\bigr].
 \label{eq:allaxesmoments}
\end{align}
Thus at most five nonidentity moments of the target are nonzero across these configurations. In testing compatibility, the measured moments that vanish must also be matched: optimizing over only the five displayed values would generally be insufficient.

\begin{proposition}[Noise-independent forbidden axes]\label{prop:supnoiseforbidden}
For the data above, $n_xn_z=0$ guarantees a $C|AB$-separable explanation for every $p,\eta\in[0,1]$, independently of $\bm a,\bm b$. For $\eta=0$ the target is $A|BC$-separable for every choice of measurement directions.
\end{proposition}
\begin{proof}
Let $\Delta_{C'}(R)=[R+C'RC']/2$, acting on $C$ only, and put $\omega=\Delta_{C'}(\widetilde\rho_{G,2})$. As a nonselective rank-one local projective measurement, $\Delta_{C'}$ produces a state separable across $C|AB$. Every added-measurement effect commutes with $C'$, so its probabilities are unchanged. The adjoint satisfies $\Delta_{C'}^\dagger(X_C)=n_x C'$. By Eq.~(\ref{eq:allaxespauli}), the only potentially altered baseline moments are
\begin{equation}
 \langle Z_BX_C\rangle_{\omega}=s n_xn_z,\qquad
 \langle X_AX_C\rangle_{\omega}=s\eta n_xn_z,
 \label{eq:allaxesshifts}
\end{equation}
where both moments vanish for the target; every other baseline moment is preserved. Hence the complete reconstructed eraser data coincide when $n_xn_z=0$. Finally $\widetilde\rho_{G,2}(0,p)$ results from $Z_A$-dephasing of the target followed by white-noise mixing, making it separable across $A|BC$.
\end{proof}

The pinched state also yields an exact noisy measurement-distance calculation. For the $X_AZ_BX_C$ readout, label the outcomes by $a,b,c\in\{-1,1\}$ and put $\kappa=n_xn_z$. The target and the pinched biseparable state have probabilities
\begin{align}
 P_{\rho}(a,b,c)&=\tfrac18(1+s\eta ab),\nonumber\\
 P_{\omega}(a,b,c)&=\tfrac18\bigl(1+s\eta ab+s\kappa bc+s\eta\kappa ac\bigr).
 \label{eq:noisybaselineprobs}
\end{align}
Consequently, for the full $X_AZ_BX_C$ joint outcome distribution,
\begin{equation}
 \operatorname{TV}(P_\rho,P_\omega)
 =\tfrac12\sum_{a,b,c}|P_\rho-P_\omega|
 =\frac{(1-p)|n_xn_z|}{2}.
 \label{eq:noisyTV}
\end{equation}
The equality uses $|b+\eta a|$ summed over $a,b$, with $0\le\eta\le1$. This is the distance to one explicit biseparable state, not the minimum distance to the entire biseparable-compatible set. At $p=0$ the distance is $|n_xn_z|/2$; at $p=1$ it vanishes.

\subsection*{Exact semidefinite characterization and attainable noisy threshold}
For the complete eraser-tomography data plus one added product-projective distribution, compatibility can be evaluated by an \emph{exact finite semidefinite program}. This reduction is specific to the measured operator space: these data do not fix independently optimized Helstrom distinguishabilities. It applies to arbitrary unit Bloch axes of $A',B',C'$ and $0\le\eta\le1$. The phrase ``exact threshold'' means that the SDP optimum coincides with the biseparable compatibility boundary for these fixed axes; it does not imply a closed analytic formula at every axis choice or optimization over the choice of axes.

For any added axes, select adapted local Pauli coordinates as follows. On $A$ take $\widetilde Z_A=X_A$ and choose $\widetilde X_A$ along the part of $A'$ perpendicular to $X_A$. On $B$ take $\widetilde Z_B=Z_B$ and $\widetilde X_B$ along the transverse part of $B'$. On $C$ take $\widetilde Z_C=X_C$ and $\widetilde X_C$ along the transverse part of $C'$. When a transverse part vanishes, choose any perpendicular axis. Complete each to a right-handed frame. These coordinate changes are local unitaries. Consequently the added axes contain only $\widetilde X$ and $\widetilde Z$, and the real baseline operator space is
\begin{equation}
 \{\id,\widetilde X_A,\widetilde Z_A\}\otimes
 \{\id,\widetilde Z_B\}\otimes\{\id,\widetilde Z_C\}.
 \label{eq:SMrealbasis}
\end{equation}
The remaining four baseline moments contain $\widetilde Y_A$ and have zero target expectations; they are part of the complete eraser-tomography data, not discarded observations. Let $O_i$ enumerate the independent traceless real measured operators, including those with zero target values, and put $d_i=\Tr[O_i\widetilde\rho_{G,2}(\eta,0)]$. There are at most 17 independent real operators. Each $O_i$ is a real linear combination of tensor products of local $\id$, $\widetilde X$, and $\widetilde Z$ operators. Here $\Gamma_j$ denotes partial transposition on qubit $j$ in the adapted basis. These Pauli factors are symmetric in the adapted computational basis, hence $O_i^{\Gamma_j}=O_i$ for \emph{every} $j=A,B,C$. By contrast, the baseline $\widetilde Y_A$ operators are antisymmetric under ordinary transposition and their target expectations are zero.

\begin{theorem}[Exact noisy compatibility with a constructive converse]\label{thm:exactnoisySDP}
Fix $0\le\eta\le1$ and arbitrary unit Bloch axes $A',B',C'$. The specified measurements comprise complete eraser tomography and one full product-projective distribution. All zero-valued moments are imposed, and no independently optimized $D_J$ values are assumed. For $\bm d\ne0$ define
\begin{align}
 s_{\mathrm c}=\max_{s,Q_A,Q_B,Q_C}\quad&s,\nonumber\\
 Q_j&=Q_j^{\mathsf T}\succeq0,\quad Q_j^{\Gamma_j}=Q_j\quad(j=A,B,C),\nonumber\\
 \Tr\sum_jQ_j&=1,\qquad \Tr(O_i\sum_jQ_j)=s d_i\quad(\forall i).
 \label{eq:SMexactSDP}
\end{align}
Here $s$ is unrestricted during the optimization. The exact physical critical noise is
\begin{equation}
 p_{\mathrm c}=\max\{0,1-s_{\mathrm c}\}.
 \label{eq:SMexactPC}
\end{equation}
The measured data admit a biseparable preparation exactly when $p\ge p_{\mathrm c}$. The optimum is attained, and a matching preparation can use at most 24 pure states, each product across one of the three bipartitions. If $\bm d=0$, the maximally mixed state already matches the data and $p_{\mathrm c}=0$.
\end{theorem}

\begin{proof}
Write any biseparable preparation as $S_A+S_B+S_C$, with each $S_j$ positive and separable across the bipartition separating qubit $j$ from the other two. First form the real symmetric matrix $R_j=(S_j+S_j^{\mathsf T})/2$. This averaging preserves all real measured moments, forces the four zero imaginary moments, and retains cut separability. Define
\begin{equation}
 Q_j=\tfrac12(R_j+R_j^{\Gamma_j}).
 \label{eq:SMsymmetrize}
\end{equation}
Since a cut-separable $R_j$ and its partial transpose are positive, $Q_j\succeq0$ and $Q_j^{\Gamma_j}=Q_j$. For each real measured observable, $O_i^{\mathsf T}=O_i^{\Gamma_j}=O_i$. Using self-adjointness of partial transposition under the trace pairing, one obtains $\Tr(O_iQ_j)=\Tr(O_iS_j)$ for every $i$ and cut $j$. Furthermore, the four other baseline moments contain one $\widetilde Y_A$ factor and vanish on the real symmetric matrices $Q_j$; their observed target values also vanish. Therefore the \emph{complete} measured distribution is preserved. Partial transposition here is only an algebraic symmetrization of cut-separable terms, not an asserted physical channel. This proves the necessary semidefinite constraints.

Conversely, $Q_j^{\Gamma_j}=Q_j$ together with $Q_j\succeq0$ implies separability across a qubit--two-qubit cut \cite{Kraus2000}; here one may also see the decomposition directly. In the cut's block basis write
\begin{equation}
 Q_j=\begin{pmatrix}D&F\\F&E\end{pmatrix},\quad
 F=F^\dagger,\quad R=E-FD^+F\succeq0.
 \label{eq:SMblockform}
\end{equation}
Positivity also forces the off-diagonal block to be supported on $D$: if $P_D=DD^+$ is the support projector, then $F=P_DFP_D$. Consequently, the generalized Schur complement $R=E-FD^+F$ is positive even when $D$ is singular.
On the support of $D$, diagonalize $(D^+)^{1/2}F(D^+)^{1/2}=\sum_k t_k|v_k\rangle\langle v_k|$ and put $f_k=D^{1/2}v_k$. Then
\begin{align}
 Q_j={}&\sum_k(|0\rangle+t_k|1\rangle)(\langle0|+t_k\langle1|)
 \otimes|f_k\rangle\langle f_k|\nonumber\\
 &+|1\rangle\langle1|\otimes R.
 \label{eq:SMproductdecomp}
\end{align}
The sum contains at most $\operatorname{rank}D\le4$ terms, and a spectral decomposition of $R$ supplies at most four more. Each $Q_j$ therefore has at most eight pure cut-product terms, hence at most 24 over the three cuts. The construction covers singular blocks via the Moore--Penrose inverse. Because the sum of the $Q_j$ is normalized and reproduces every real measured moment while its remaining imaginary baseline moments vanish, undoing the local coordinates gives an exact biseparable match to the full measured data.

At $s=0$, $Q_j=\id/24$ is strictly positive on the symmetry subspace; compactness bounds and attains the optimum. Mixing a maximizing preparation $\omega_{\mathrm c}$ with $\id/8$ yields, for $0\le s\le\min\{1,s_{\mathrm c}\}$,
\begin{equation}
 \omega(s)=\frac{s}{s_{\mathrm c}}\,\omega_{\mathrm c}+
 \left(1-\frac{s}{s_{\mathrm c}}\right)\frac{\id}{8}.
 \label{eq:SMattainnoisy}
\end{equation}
This is a physical convex mixture matching the data at white-noise fraction $p=1-s$ and proves attainment throughout the compatible range. The zero-data case is immediate.
\end{proof}

The same symmetrization proves that the usual positive-partial-transpose (PPT) mixture relaxation \cite{Jungnitsch2011} is \emph{exact for compatibility with these measured data}: applying the construction to positive PPT components produces cut-separable alternatives with the same observations. It does not assert that arbitrary three-qubit PPT mixtures are biseparable.

The dual is finite as well. Write $H=\sum_i\lambda_i O_i$, with $\sum_i\lambda_i d_i=1$. For each cut there is a real positive matrix $P_j$ such that
\begin{equation}
 s_{\mathrm c}=\min_{\beta,\boldsymbol\lambda,P_j}\ \beta,\qquad
 \beta\id-H=\tfrac12(P_j+P_j^{\Gamma_j})\quad(j=A,B,C),\quad P_j=P_j^{\mathsf T}\succeq0.
 \label{eq:SMexactSDPdual}
\end{equation}
The strictly feasible primal point yields strong duality. Each dual feasible point gives an accessible biseparable witness, and the optimizing bound matches the primal preparation endpoint. All constraints are linear matrix inequalities; this is an exact optimization characterization, although numerical solution of the SDP alone does not certify its last decimal digits.

For the rational axes in Eq.~(\ref{eq:SMrationalaxes}) below, the exact rational certificate independently verifies all 17 real moment equations, the three partial-transpose symmetries, the three dual identities, and positivity through 48 rational $LDL^{\mathsf T}$ pivots. It gives
\begin{align}
 A'&=(3X_A+4Z_A)/5,\quad B'=(4X_B+3Z_B)/5,\nonumber\\
 C'&=(4X_C+3Z_C)/5,\quad \eta=1,\label{eq:SMrationalaxes}\\
 0.14919347&\le p_{\mathrm c}\le0.14919348.
 \label{eq:SMrationalbracket}
\end{align}
The upper endpoint is realized by rational feasible matrices and an explicit 24-term cut-product decomposition. A matching rational dual certificate excludes biseparable data below the lower endpoint. Numerical optimization gives $p_{\mathrm c}\simeq0.1491934751$; the rigorous statement is the bracket, not an exact decimal equality. The optional decimal 24-term decomposition is a separate numerical consistency check.

The GHZ-family biseparability boundary gives $0\le p_{\rm c}\le\eta/(\eta+3/4)$ for the three-qubit target. For the specified data, the rational-axis certificate in Eq.~(\ref{eq:SMrationalbracket}) provides a sharper interval.

For $\eta=1$ and $n_xn_z\ne0$, the additional axes of the pure-state theorem produce data outside the compact biseparable-compatible set at $p=0$. Continuity along the white-noise line therefore gives $p_{\rm c}>0$ for that choice of the other two axes. This establishes strict but not uniform noise tolerance, without providing an explicit fraction.

\subsection*{Statistical separation from locally dephased comparators}
\paragraph{Noiseless measurement comparison.}
The argument extends to all $m\ge2$ without assuming a globally pure competing state. Consider the target $\rho_{G,m}(\eta)$ without white-noise mixing and an added product-projective context containing the rank-one effects of $\bm n\cdot\bm\sigma_j$ on one fixed marker $j\ge2$. Define the biseparable state $\omega_j=\Delta_{\bm n,j}(\rho_{G,m}(\eta))$, where the rank-one pinching is $\Delta_{\bm n,j}(R)=[R+(\bm n\cdot\bm\sigma_j)R(\bm n\cdot\bm\sigma_j)]/2$. Every added context effect is fixed in the Heisenberg picture, regardless of the other parties' measurement bases, so its entire outcome distribution is identical for $\rho_G$ and $\omega_j$.

Conditioning on the baseline $Z_1$ outcome $b$ puts the path qubit in $[\id_A+\eta(-1)^b X_A]/2$ and every other marker in $|b\rangle$. Hence the full baseline probabilities, with $x_\ell=\pm1$ the outcomes of $X_\ell$ for $\ell\ge2$, obey
\begin{equation}
 P_G(b,\bm x)=2^{-m},\qquad
 P_{\omega_j}(b,\bm x)=2^{-m}\left[1+(-1)^b x_j\kappa\right],\quad \kappa=n_xn_z.
 \label{eq:universalpinchprobs}
\end{equation}
The conditional path density matrix for fixed $b,\bm x$ is the same normalized path state in the two preparations. Thus additional path tomography cannot distinguish them beyond the displayed marker-outcome factor. Their total variation is $|\kappa|/2$. Taking the Kullback--Leibler (KL) divergence with $P_G$ as the first argument yields
\begin{equation}
 D_{\rm KL}(P_G\Vert P_{\omega_j})=
 -\frac12\ln(1-\kappa^2), \qquad |\kappa|\le\tfrac12.
 \label{eq:universalpinchKL}
\end{equation}
For this comparator, $D_{\rm KL}=\kappa^2/2+O(\kappa^4)$ as $\kappa\to0$. Distinguishing these two distributions at fixed nontrivial error probabilities therefore requires at least order $\kappa^{-2}$ independent baseline trials.

\paragraph{Noisy baseline readouts.}
For the noisy $C|AB$-separable comparator in Eq.~(\ref{eq:noisybaselineprobs}), set $s=1-p$ and $\kappa=n_xn_z$. For the three full baseline distributions, let $D_{\mathrm{KL}}^{(X)},D_{\mathrm{KL}}^{(Y)},D_{\mathrm{KL}}^{(Z)}$ be the directed Kullback--Leibler divergences from the target to this comparator in the $X_AZ_BX_C$, $Y_AZ_BX_C$, and $Z_AZ_BX_C$ settings, respectively. With
\begin{equation}
 w_t=\frac{1+s\eta t}{2},\qquad
 \gamma_t=\frac{s\kappa(1+\eta t)}{1+s\eta t},\qquad t=\pm1,
 \label{eq:SMnoisygamma}
\end{equation}
where zero-weight terms are interpreted by continuity, direct summation of the eight-outcome probabilities yields
\begin{equation}
 D_{\mathrm{KL}}^{(X)}=-\frac12\sum_{t=\pm1}w_t\ln(1-\gamma_t^2),\qquad
 D_{\mathrm{KL}}^{(Y)}=D_{\mathrm{KL}}^{(Z)}=-\frac12\ln(1-s^2\kappa^2).
 \label{eq:SMnoisyKL}
\end{equation}
These divergences satisfy
\begin{equation}
 D_{\mathrm{KL}}^{(X)}\ge D_{\mathrm{KL}}^{(Y)}=D_{\mathrm{KL}}^{(Z)}.
 \label{eq:SMKLorder}
\end{equation}
Indeed, Jensen's inequality applied to $f(z)=-\ln(1-z)$ gives $D_{\mathrm{KL}}^{(X)}\ge -\tfrac12\ln(1-\sum_t w_t\gamma_t^2)$, and
\begin{equation}
 \sum_t w_t\gamma_t^2
 =s^2\kappa^2\frac{1+(1-2s)\eta^2}{1-s^2\eta^2}\ge s^2\kappa^2,
 \label{eq:SMKLJensen}
\end{equation}
for $s\eta<1$, with endpoint cases taken by continuity.
For $\kappa\to0$ at fixed $s\eta<1$,
\begin{equation}
 D_{\mathrm{KL}}^{(X)}=\frac{s^2\kappa^2}{2}\frac{1+(1-2s)\eta^2}{1-s^2\eta^2}+O(\kappa^4).
 \label{eq:SMnoisyKLexp}
\end{equation}
Thus the statistical distinction between the target and this dephased comparator diminishes near the excluded marker directions.

\subsection*{General distance bound for local dephasing}
For a collection of local measurements along $\bm n_k$, compare a state with the result of dephasing one qubit along $\bm u$. The following bound applies to each measured setting, irrespective of the state on the other parties. For measured local directions $\bm n_k$ and a fixed dephasing axis $\bm u$, the effect difference is $E_r^{(k)}-\Delta_{\bm u}(E_r^{(k)})=r\bm n_{k,\perp}\cdot\bm\sigma/2$, where $r=\pm1$ is the local measurement outcome. Against any complementary POVM $F_y$,
\begin{equation}
 P_\rho(r,y|k)-P_{\Delta_{\bm u}(\rho)}(r,y|k)=\tfrac r2\Tr[\rho(\bm n_{k,\perp}\cdot\bm\sigma\otimes F_y)].
\end{equation}
Write $\operatorname{TV}$ for total variation distance. Writing $h_k=\|\bm n_{k,\perp}\|$, we have $-h_k\id\le\bm n_{k,\perp}\cdot\bm\sigma\le h_k\id$; summing over the local outcomes gives
\begin{equation}
 \operatorname{TV}(P_\rho^{(k)},P_{\Delta_{\bm u}(\rho)}^{(k)})\le\tfrac12\sqrt{1-(\bm n_k\cdot\bm u)^2}.
\end{equation}
A product input with local Bloch vector parallel to $\bm n_{k,\perp}$ attains this bound. For a collection of measured axes, the best uniform bound supplied by this construction is
\begin{equation}
 \epsilon_{\rm pinch}=\tfrac12\inf_{\|\bm u\|=1}\max_k\sqrt{1-(\bm n_k\cdot\bm u)^2}.
 \label{eq:generalpinchminimax}
\end{equation}
For two unoriented axes at acute angle $\alpha$, minimizing the maximum over both axes gives $\tfrac12\sin(\alpha/2)$, attained at the bisector: equivalently maximize $\min_k|\bm n_k\cdot\bm u|=\cos(\alpha/2)$. This minimax statement is restricted to single-site pinching channels, not all biseparable simulators or a statistical lower bound.

\section{A fixed tilted witness and its biseparable bound}\label{sec:supp:7}
The exact SDP uses all the specified measurement statistics. For comparison, the main article introduces a single fixed witness that can be evaluated from the same tilted axes. Here we give its coefficients, prove the cutwise rational bound, and compute its response on the noisy target.

The eraser tomography provides the $Z_BX_C$ product outcome distribution together with the $X_A$ and $Z_A$ path readouts. Thus its accessible Pauli products are $I$, $Z_B$, $X_C$, $Z_BX_C$, and their products with $X_A$ or $Z_A$. A single further product configuration, $T_AT_BT_C$, supplies the tilted terms and their marginals. We evaluate a fixed rational score in the span of these measured correlators. The coefficient list below specifies the score completely; the global optimization proved here concerns states for this fixed operator, not a search over all witness coefficients or measurement axes. Let $T_A=(3X_A+4Z_A)/5$, $T_B=(4X_B+3Z_B)/5$ and $T_C=(4X_C+3Z_C)/5$. Suppressing identities on the $BC$ subsystem, put
\begin{align}
 10000H_T={}&\id_A\otimes\bigl[-1029X_C+1948Z_B-2035Z_BX_C
 +2741T_C-3786T_B+3544T_BT_C\bigr]\nonumber\\
 &+X_A\otimes\bigl[-2599\id-2797X_C+4229Z_B-1605Z_BX_C\bigr]\nonumber\\
 &+Z_A\otimes\bigl[-2850\id-3399X_C-1043Z_B-1844Z_BX_C\bigr]\nonumber\\
 &+T_A\otimes\bigl[7315T_C-2052T_B+5080T_BT_C\bigr].
 \label{eq:newH}
\end{align}
Every coefficient multiplies an observable that is obtainable from the complete $Z_BX_C$ eraser/path-tomography data or from the single added $T_AT_BT_C$ outcome distribution. The score uses no marker setting beyond the specified tilted configuration. It uses the $X_A$ and $Z_A$ baseline tomography readouts separately; these are already assumed included in the complete $\tau$ matrices. The operator is real, Hermitian, and traceless.

\subsection*{Reduction to product states across each cut}
For each bipartition $A|BC$, $B|AC$, or $C|AB$, write $H_T$ in the product form and define real symmetric matrices on the two-qubit complement,
\begin{equation}
 H_\nu^{(j)}=\tfrac12\Tr_j[(\sigma_\nu^{(j)}\otimes\id_{\overline j})H_T],\quad \nu=0,x,z,
 \label{eq:newpencils}
\end{equation}
where $\sigma_0=\id$, and the product factors are reordered appropriately before tracing. There is no isolated-site $Y$ term. For a product across the cut with isolated-site Bloch coordinates $(x,y,z)$, its optimal two-qubit complement yields the largest eigenvalue of
\begin{equation}
 \mathcal H_j(x,z)=H_0^{(j)}+xH_x^{(j)}+zH_z^{(j)},\quad x^2+z^2\le1.
 \label{eq:newpencil}
\end{equation}
Convexity of the largest eigenvalue shows that the maximum occurs on the unit circle.

\subsection*{Global biseparable bound for the fixed witness}
For the fixed score $H_T$, we first establish a rational upper bound (distinct from its exact biseparable maximum),
\begin{equation}
 \Tr(H_T\rho)\le\beta_T:=\frac{4257}{5000}=0.8514
 \qquad(\rho\in\Bisep).
 \label{eq:sharperbound}
\end{equation}
Parametrize the unit circle by $x=2t/(1+t^2)$ and
$z=(1-t^2)/(1+t^2)$. For each bipartition define
\begin{equation}
 M_j(t)=(\beta_T\id-H_0^{(j)})(1+t^2)
 -2tH_x^{(j)}-(1-t^2)H_z^{(j)}.
 \label{eq:sturmM}
\end{equation}
After clearing positive denominators, the leading principal minors of
orders $k=1,2,3,4$ are polynomials of degree $2k$. For each cut and
order, their Sturm variations satisfy $V(-\infty)=V(+\infty)=k$.
By Sturm's theorem, the difference of these variations counts distinct real zeros, and here it vanishes. Thus none of the twelve minors has a real zero. All are positive at
$t=0$ and hence throughout the real line. Their values are also positive
at the omitted point $(x,z)=(0,-1)$. Sylvester's criterion proves
$\beta_T\id-\mathcal H_j(x,z)>0$ on the circle. Convexity of the largest
eigenvalue extends Eq.~(\ref{eq:sharperbound}) to the Bloch disk and to
all biseparable mixtures.

The fixed-witness bound is sufficient; the compatible-state boundary is obtained separately in Sec.~\ref{sec:supp:6}.

\subsection*{Target response and confidence bounds}
For $\widetilde\rho_{G,2}(\eta,p)$, direct Pauli expansion gives
\begin{equation}
 \Tr[H_T\widetilde\rho_{G,2}(\eta,p)]=(1-p)\left(\frac{2423}{6250}+\frac{38273\eta}{62500}\right).
 \label{eq:newresponse}
\end{equation}
The certified violation requires
\begin{equation}
 p<1-\frac{4257/5000}{2423/6250+38273\eta/62500},
 \label{eq:newnoise}
\end{equation}
when the right side is positive. At $\eta=1$, this is $p<18581/125006\approx0.148641$. At $p=0$ it holds when $\eta>57965/76546\approx0.757257$. These are guaranteed witness-detection regions, not the exact noisy compatibility frontier or the GHZ-family GME boundary.

\section{Finite-trial certification and measurement requirements}\label{sec:supp:8}
The main article, Sec.~V, gives the finite-trial test and its interpretation. Here we state the score ranges, derive concentration and allocation bounds, compare measurement costs, and describe calibration and accepted-trial assumptions. We assume trusted qubit measurements, independent and identically distributed preparations from a setting-independent source, and predetermined positive readout counts. Within-trial multipartite outcomes may be correlated. Temporal memory or drift requires a different concentration analysis; unverified marginal calibration does not establish coverage in that setting.

\subsection*{A single concentration bound for the combined score}
To form independent single-trial scores, allocate half of the $A$-identity baseline terms in Eq.~(\ref{eq:newH}) to each of the $X_A$ and $Z_A$ readout blocks; allocate their corresponding $X_A$ or $Z_A$ terms entirely to those blocks. The tilted terms form the third block.

For the fixed tilted witness, the score ranges in the $X_AZ_BX_C$, $Z_AZ_BX_C$, and $T_AT_BT_C$ contexts are, respectively,
\begin{equation}
 R_X=\frac{17029}{10000},\qquad
 R_Z=\frac{18272}{10000},\qquad
 R_T=\frac{467}{125}.
 \label{eq:SIranges}
\end{equation}
For context $k$, let each trial score $S_{ki}$ lie in an interval of range $R_k$. The estimator is $\widehat{\mathcal Q}=\sum_kN_k^{-1}\sum_{i=1}^{N_k}S_{ki}$. The independent summands have ranges $R_k/N_k$, so Hoeffding's theorem \cite{Hoeffding1963} yields
\begin{equation}
 \Pr(\widehat{\mathcal Q}-\mathcal Q\ge t)
 \le \exp\!\left[-2t^2\left/\sum_k\frac{R_k^2}{N_k}\right.\right].
 \label{eq:SIcombined}
\end{equation}
For a null bound $\mathcal Q\le b$, reject the null when $\widehat{\mathcal Q}>b+c_\alpha$, where
\begin{equation}
 c_\alpha=\sqrt{\frac{\ln(1/\alpha)}2\sum_k\frac{R_k^2}{N_k}}.
 \label{eq:SImargin}
\end{equation}
The type-I error is at most $\alpha$. For target mean $b+\Delta$, the lower-tail bound gives power at least $1-\beta$ if $c_\alpha+c_\beta<\Delta$. At fixed $N_{\rm tot}=\sum_kN_k$,
\begin{equation}
 \left(\sum_k\frac{R_k^2}{N_k}\right)N_{\rm tot}
 \ge\left(\sum_kR_k\right)^2,
 \qquad N_k:N_\ell=R_k:R_\ell.
 \label{eq:SIrangeallocation}
\end{equation}
Equality is attained for continuous allocations proportional to the ranges; rounded counts must be rechecked. This optimality concerns the stated Hoeffding bound, not all statistical tests. For a target excess $\Delta>0$, false-positive allowance $\alpha$, and missed-detection allowance $\beta$, this gives the sufficient continuous total
\begin{equation}
 N_{\rm tot}>
 \frac{[\sqrt{\ln(1/\alpha)}+\sqrt{\ln(1/\beta)}]^2}{2\Delta^2}
 \biggl(\sum_{k=X,Z,T}R_k\biggr)^2.
 \label{eq:SItotal}
\end{equation}
At the ideal target with $\alpha=\beta=0.05$, the explicit integer allocation $(N_X,N_Z,N_T)=(3356,3601,7361)$ meets the strict criterion and totals $14\,318$ accepted trials.

By comparison, simultaneous two-sided context intervals with total failure probability $\delta$ give
\begin{equation}
 c_{\rm union}=\sum_{k=1}^{K}
 R_k\sqrt{\frac{\ln(2K/\delta)}{2N_k}}.
 \label{eq:SIunion}
\end{equation}
At fixed total, minimizing this separate-context expression gives $N_k\propto R_k^{2/3}$. When the false-positive and missed-detection allowances are both $0.05$, the sufficient target-power condition doubles the corresponding single-margin correction. For the ideal tilted score, the combined and union sufficient totals are $14315.865$ and $65553.124$, respectively. The ratio is approximately $4.579$. This is a comparison of sufficient bounds for the same witness, measurements and error goals; it neither minimizes the hypothesis-testing risk nor improves on the standard two-Pauli protocol. The required full eraser-tomography context with a $Y_A$ readout, and any independently determined $D_J$, are not included in these accepted witness-score counts.

\begin{figure}[htbp]
 \centering
 \includegraphics[width=0.94\linewidth]{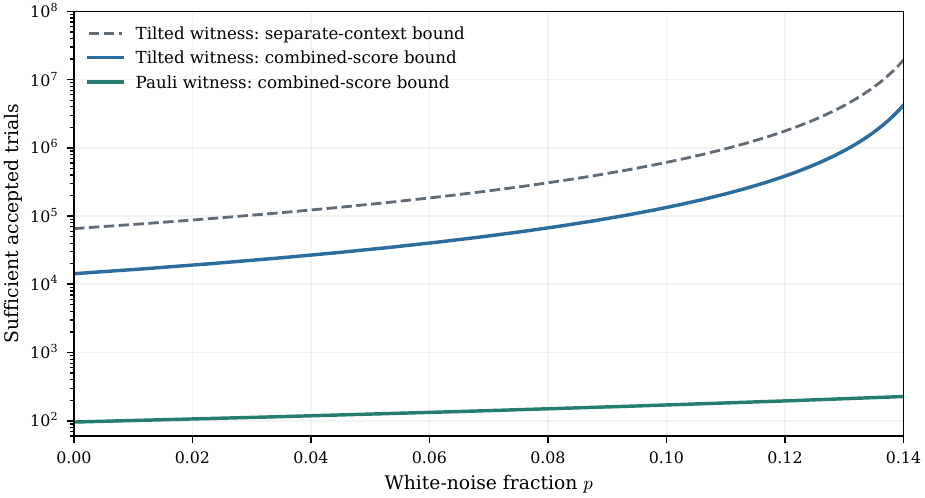}
 \caption{Sufficient accepted-trial counts for the three-qubit target at $\eta=1$ with type-I error at most $0.05$ and target detection probability at least $0.95$. Curves are continuous sufficient totals before rounding. Pooling the tilted-witness score tightens its Hoeffding bound compared with separate-context intervals, but the conventional Pauli test remains less costly under matched ideal assumptions. These counts exclude calibration, losses, and independent distinguishability measurements.}
 \label{fig:supp:trials}
\end{figure}

\subsection*{Two Pauli pass events}
The positive-facet test uses $\widehat u+\widehat v$ with two unit-range Bernoulli scores. Its combined margin is
\begin{equation}
 c_\alpha=\sqrt{\frac{\ln(1/\alpha)}2\left(\frac1{N_X}+\frac1{N_Z}\right)}.
\end{equation}
The negative facet uses $\widehat v-\widehat u$ with the same ranges. A facet is chosen before testing; if either facet is tested, divide the total false-positive allowance between them. At $m=2,\eta=0.8,p=0.05$, the means are $u=0.975,v=0.8675$, with excess $\Delta=0.3425$. For $N_X=N_Z=512$ and $\alpha=\beta=0.05$, the combined margin is approximately $0.07649$, so twice this margin is below the mean excess. The result is a conservative ideal-model power guarantee, not measured laboratory efficiency.

For ideal independent Bernoulli events, the variance is
\begin{equation}
 \operatorname{Var}(\widehat u+\widehat v)
 =\frac{u(1-u)}{N_X}+\frac{v(1-v)}{N_Z}.
 \label{eq:bernoullivar}
\end{equation}
Let $a=u(1-u)$ and $b=v(1-v)$. At fixed $N=N_X+N_Z$ with a,b positive, Cauchy--Schwarz gives $N(a/N_X+b/N_Z)\ge(\sqrt a+\sqrt b)^2$, with equality when
\begin{equation}
 N_X:N_Z=\sqrt{u(1-u)}:\sqrt{v(1-v)}.
 \label{eq:allocation}
\end{equation}
This allocation assumes known planning probabilities and continuous counts; actual counts are rounded, and all settings remain necessary for a general-state test. The distribution-free certificate instead uses the declared fixed counts and the combined-score Hoeffding bound.

\subsection*{Systematic errors and accepted trials}
With outcome-distribution biases bounded by $e_X,e_Z$, the corrected positive-facet condition is
\begin{equation}
 \widehat u+\widehat v>\tfrac32+c_\alpha+e_X+e_Z.
 \label{eq:paulicalibration}
\end{equation}
For the tilted score the allowance is $\sum_kR_ke_k$, as in the main-text test. Writing $B=\sum_kR_ke_k$, a sufficient target-power condition is $c_\alpha+c_\beta+2B<\Delta$ when $\Delta$ is the ideal mean excess: one $B$ covers the downward mean shift and the other the threshold correction. For the Pauli sum, $B=e_X+e_Z$. A marginal recording-error bound $\epsilon_{kj}$ for party $j$ in context $k$, justified independently of the witness samples, gives the conservative distributional error $e_k=\min(1,\sum_j\epsilon_{kj})$ by coupling and the union bound; no independence of readout errors within a trial is needed. A separately calibrated local axis deviation $\theta_{kj}$ contributes at most $\sin(\theta_{kj}/2)$ in total variation, so $\sum_j\sin(\theta_{kj}/2)$ can be conservatively added to $e_k$. Recording and axis errors are distinct mechanisms, and the calibration must remain applicable to the analyzed trials.

Under independent heralded state- and setting-independent survival probability $\zeta$ per party, the mean launch cost is $N/\zeta^{m+1}$ for $N$ accepted trials. This does not cover uncontrolled coincidence postselection. Total-resource comparisons also count eraser tomography, discrimination, preparation and calibration. Supplemental Fig.~\ref{fig:supp:trials} compares accepted witness trials under common ideal assumptions.

\subsection*{Measurement acquisition and numerical cross-checks}
In the baseline configuration $Z_1X_2\cdots X_m$, the $X_A$ and $Y_A$ readouts provide the two conditional path-coherence quadratures, while $Z_A$ supplies conditional path populations. Together they determine each unnormalized matrix $\tau_t$ without discarding marker outcomes. The full $Z_AX_1\cdots X_m$ distribution yields the parity pass probability; $X_AZ_1\cdots Z_m$ yields the simultaneous-correlation pass probability. Independent optimal distinguishabilities $D_J$ require separate measurements and are not provided by these readouts automatically. For the three-qubit fixed witness, the baseline $X_AZ_BX_C$ and $Z_AZ_BX_C$ contexts plus the tilted $T_AT_BT_C$ context suffice to evaluate the score; the complete eraser data additionally requires $Y_AZ_BX_C$ unless the phase quadrature is already known. All context counts refer to trusted settings and accepted trials rather than total experimental launches.